%% file: main.tex
\documentclass[a4paper,USenglish]{lipics-v2021}

\pdfoutput=1
\hideLIPIcs

\usepackage{algorithm}
\usepackage{algpseudocode}
\usepackage{smartalgorithmic}
\usepackage{xcolor}

\usepackage{graphicx}
\usepackage{tikz}
\usetikzlibrary{arrows.meta,positioning,fit,calc,shapes.misc,shapes.geometric,decorations.pathmorphing,patterns.meta,shadows.blur}
\usepackage{mathtools}
\usepackage{amssymb}
\usepackage{amstext}
\usepackage{amsmath}
\usepackage{amsthm}
\usepackage{amsfonts}
\usepackage{bm}

\input{definitions}

\title{A Deterministic Commitment Rule over Unstructured DAGs for Optimistic Replication}

\titlerunning{Deterministic Commitment over Unstructured DAGs}

\author{João Paulo {Bezerra}}{Université Paris-Saclay, CEA, LIST, Palaiseau, France}{Joao-Paulo.BEZERRADEARAUJO@cea.fr}{https://orcid.org/0000-0003-3620-899X}{}

\author{Petr {Kuznetsov}}{Télécom Paris, Institut Polytechnique de Paris, Palaiseau, France}{petr.kuznetsov@telecom-paris.fr}{https://orcid.org/0000-0003-1148-1228}{}

\author{Léo {Olivier}}{Université Paris-Saclay, CEA, LIST, Palaiseau, France}{leo.olivier@cea.fr}{https://orcid.org/0000-0001-9815-0530}{}

\author{Sara {Tucci-Piergiovanni}}{Université Paris-Saclay, CEA, LIST, Palaiseau, France}{sara.tucci@cea.fr}{https://orcid.org/0000-0001-9738-9021}{}

\authorrunning{J. P. Bezerra, P. Kuznetsov, L. Olivier, S. Tucci-Piergiovanni}

\Copyright{J. P. Bezerra, P. Kuznetsov, L. Olivier, S. Tucci-Piergiovanni}



\category{}

\relatedversion{}

\nolinenumbers

\EventEditors{John Q. Open and Joan R. Access}
\EventNoEds{2}
\EventLongTitle{42nd Conference on Very Important Topics (CVIT 2016)}
\EventShortTitle{CVIT 2016}
\EventAcronym{CVIT}
\EventYear{2016}
\EventDate{December 24--27, 2016}
\EventLocation{Little Whinging, United Kingdom}
\EventLogo{}
\SeriesVolume{42}
\ArticleNo{23}

\begin{document}

\raggedbottom

\maketitle

\input{definitions_tickz}

\begin{abstract}
In optimistic replication protocols, 
each replica uses the local state to execute requests from its clients in a \emph{wait-free} manner, without synchronizing with other replicas.
The outcomes may be revised later, making optimistic protocols only \emph{eventually} consistent. 
To compensate for this consistency loss, an additional \emph{commitment} mechanism can be employed: an underlying consensus protocol that allows replicas to periodically agree on the evolution of the object state.

In this paper, we introduce a simple framework for optimistic replication based on a \emph{causality DAG}: every replica represents the object state via a causally ordered set of client operations.
The framework incorporates a \emph{deterministic} commitment rule that incurs virtually zero message overhead beyond the basic dissemination of submitted operations.
\end{abstract}

\section{Introduction}
\label{sec:intro}

\subparagraph{Optimistic replication.}
Distributed systems face a fundamental tradeoff between responsiveness and consistency. Strongly consistent protocols require replicas to coordinate before responding to client requests, which can significantly increase latency.
Optimistic replication takes the opposite approach: replicas execute client operations immediately using their local state and perform synchronization in the background. As a result, clients obtain low-latency responses, while replicas gradually converge towards a common state~\cite{saito2005optimistic}.

The behavior of optimistic protocols is commonly captured through eventual consistency. Replicas may temporarily disagree on the order in which operations should be applied, causing local states to diverge.
Nevertheless, after exchanging sufficient information, correct replicas eventually converge. In many applications, it is desirable to complement eventual consistency with a commitment mechanism that allows replicas to irrevocably determine a common ordering of operations.
Such a mechanism enables strongly consistent operations, safe garbage collection, and reconfiguration.

Conflict-Free Replicated Data Type (CRDT)~\cite{shapiro2011conflict} protocols are among the most widely used optimistic replication techniques.
CRDT protocols satisfy strong eventual consistency: whenever replicas observe the same set of operations, they reach the same state.
Moreover, they tolerate an arbitrary number of failures and avoid the need for consensus during normal operation.
However, CRDTs do not naturally provide a commitment mechanism.

Numerous CRDT implementations rely on causal broadcast~\cite{burckhardt2014replicated,almeida2024approaches,waitfree2025} to disseminate operations.
In particular, Baquero et al.~\cite{baquero2017pure} and Almeida~\cite{almeida2024approaches} describe a generic framework based on a \emph{partially ordered log}, where the replicated state is derived by locally interpreting a causally ordered set of operations.
Following this approach, we view the causality relation between disseminated operations as a directed acyclic graph (DAG) and show that this structure can be leveraged not only to guarantee convergence but also to derive a deterministic commitment mechanism.

\subparagraph{Deterministic commitment from the DAG.}
DAG-based protocols have recently attracted considerable attention due to their simplicity and high throughput.
Most existing DAG-based solutions operate in the Byzantine setting and target blockchain applications, whereas we consider crash failures, the conventional failure model for optimistic replication systems.
%
%
These blockchain solutions typically construct DAGs with \emph{layered structures} by organizing vertices into asynchronous rounds~\cite{keidar2021all,danezis2022narwhal} and requiring replicas to collect sufficient information before issuing a new vertex.
Commitment is obtained through additional mechanisms such as randomization~\cite{moser1999byzantine,baird2016swirlds,gkagol2019aleph} or waiting for designated leaders~\cite{keidar2023cordial,babel2023mysticeti}.

Rather than constructing a layered DAG, we consider the causality DAG that naturally emerges from the exchange of replica messages in an optimistic replication protocol, similarly to the partially ordered logs used in CRDT systems.
Since replicas do not wait for any particular message before issuing new vertices, the DAG remains \emph{unstructured}.
%
%
To ensure progress, replicas equip every disseminated message with the output of an eventual leader oracle $\Omega$, which encapsulates the smallest amount of synchrony for establishing a total order on the inputs~\cite{chandra1996weakest}. 
Therefore, the DAG captures not only causality relations between operations, but also information about the underlying system synchrony.
Using this information, replicas can locally and deterministically derive a common total order whenever a majority of replicas remain correct and $\Omega$ eventually stabilizes on the same leader at all correct replicas.
No rounds, randomization, or separate ordering protocol are required.

\subparagraph{Contributions.} To summarize, our contributions are threefold:
\begin{itemize}
    \item We show how the information produced by an eventual leader oracle can be embedded into a causality DAG, allowing replicas to derive deterministic commitment without rounds, randomization, or timeout-driven coordination.
    \item We use the resulting framework to construct a generic replication protocol supporting both optimistic and strong consistency criteria.
    \item We devise an efficient commitment mechanism that incurs virtually no communication overhead\footnote{Efficient implementations of $\Omega$ can be realized with minimal communication overhead, requiring replicas to exchange only a few lightweight heartbeat messages~\cite{aguilera2001stable,schiper2008robust}.} beyond the dissemination layer already used by many optimistic replication protocols.
\end{itemize}

\subparagraph{Roadmap.} The rest of the paper is organized as follows.
We discuss related work in Section~\ref{sec:relWork}.
In Section~\ref{sec:model}, we present our system model and specify the optimistic replication problem.
We present the commitment framework and prove its correctness in Section~\ref{sec:DAGframe}.
We give our concluding remarks in Section~\ref{sec:conclusion}.

\section{Related Work}
\label{sec:relWork}

\subparagraph{Optimistic Replication.}
The idea of optimistic replication~\cite{saito2005optimistic} is to prioritize availability over consistency.
The local state of the shared data is represented by a set of submitted updates, and the clients' operations are applied to the state without synchronizing with other replicas. 
Consistency is therefore only \emph{eventual}: during periods in which no new updates are applied, replicas converge to the same state.

This approach has found numerous practical applications~\cite{vogels-ec,cassandra,bayou,dynamo}, including many implementations of \emph{CRDTs}~\cite{shapiro2011conflict}.
In the optimistic setting, message causality has been used since early work to identify and resolve conflicts among concurrent operations~\cite{ladin1992providing,lloyd2011don}.
In the Pure Operation-Based CRDT framework introduced by Baquero et al.~\cite{baquero2017pure} and later advocated by Almeida~\cite{almeida2024approaches}, replicas maintain a partially ordered log of operations together with their causal dependencies.
The state of the replicated object is obtained by locally interpreting this partially ordered structure according to the object semantics. However, operations are applied only optimistically, and replicas may never learn whether certain operations have become irreversible.

Mixed-consistency systems, by contrast, seek to provide stronger guarantees for selected critical operations, typically through a commitment mechanism that enables strongly consistent execution~\cite{guerraoui2016incremental}.
In previous approaches, however, adding a commitment mechanism on top of an optimistic service incurs significant costs: strongly consistent operations may lack fault tolerance~\cite{li2012making}, block progress~\cite{li2018fine}, or require an external atomic broadcast protocol~\cite{kokocinski2021mixing,wojciechowski2026creek}, thereby increasing communication overhead.
In contrast, our approach relies solely on the dissemination layer (causal broadcast), together with a lightweight leader oracle, to both commit operations and derive their optimistic ordering.

Closer to our work, Dubois et al.~\cite{dubois2015weakest} present an eventual total-order broadcast protocol in which replicas maintain a DAG capturing causal dependencies, while the replica currently elected by $\Omega$ periodically disseminates a causality-preserving total order.
The resulting order eventually stabilizes, but replicas cannot determine when this has occurred, since the stabilization time of $\Omega$ is unknown.
In contrast, our commitment rule allows every replica to locally determine a committed prefix whose order is guaranteed to remain stable.

\subparagraph{DAG-based protocols.}
DAG-based protocols have received significant attention in the design of high-throughput blockchain and state machine replication systems~\cite{baird2016swirlds,gkagol2019aleph,keidar2021all,danezis2022narwhal,spiegelman2022bullshark,keidar2023cordial,amores2024we,jovanovic2025mahi}.
A common design consists of organizing the DAG into asynchronous rounds~\cite{keidar2021all}, where replicas disseminate blocks and wait for a quorum of blocks before progressing.
Earlier works also explored less structured DAG constructions~\cite{moser1999byzantine,baird2016swirlds}.

To derive commitment, prior protocols typically rely on mechanisms such as external leader election~\cite{danezis2022narwhal}, common coins or randomized voting~\cite{gkagol2019aleph,danezis2022narwhal,keidar2023cordial}, timeout-based leader proposals~\cite{keidar2023cordial,babel2023mysticeti}, or assumptions on randomized network schedulers~\cite{moser1999byzantine,jovanovic2025mahi}.
In contrast, commitment in our DAG is obtained solely by interpreting the DAG and information carried by the vertices.
Information provided by the eventual leader oracle $\Omega$ is embedded directly into disseminated messages, allowing replicas to commit without waiting for quorum certificates, or executing an auxiliary leader-election or common-coin protocol.

\section{System Model and Problem Statement}
\label{sec:model}

\subparagraph{Processes, failures, and communication.} The system is composed of a fixed set of processes (or replicas) $\Pi$, each of which is a sequential automaton assigned a deterministic algorithm.
A distributed protocol is an array of algorithms $\cA$, one for each process.
Processes communicate by exchanging messages.
In each algorithm step $e$, a single process receives an input (a message, an application call or a special null value $\lambda$), performs some local computation based on the input and its current state, transitions to a new state and produces an output (a set potentially including messages, an application response, and the null value $\lambda$).
A \emph{run} (or \emph{execution}) $e_1e_2\ldots$ is a sequence of algorithm steps (or \emph{events}).
We assume a \emph{crash failure model}: processes may stop taking steps but always follow their assigned algorithm. In an infinite execution, a process is \emph{correct} if it takes an infinite number of steps, and it is \emph{faulty} otherwise.
In every run, at least a \emph{majority} of processes is correct.\footnote{This assumption is only required for the commitment of messages, as it is not necessary for optimistic progress, which relies on the liveness of the underlying causal dissemination mechanism.}

There is no upper bound on message delays.
Each message sent in a run is \emph{unique} and is received \emph{at most once}. Moreover, if there is an event $e$ in which a process $q$ receives a message $m$ from a process $p$, then there is an event $e'$ preceding $e$ in which $p$ sends $m$ to $q$.

\subparagraph{Happens-before and causality.} For a given execution, an event $e$ \emph{happens-before}~\cite{lamport1978time,cachin2011introduction} another event $e'$ if any of the following conditions is satisfied:

\begin{enumerate}
    \item $e$ and $e'$ occur at the same process $p$, and $p$ executes step $e$ before $e'$.
    \item A message $m$ is sent in $e$ and received in $e'$.
    \item There exists an event $e''$ such that $e$ happens-before $e''$ and $e''$ happens-before $e'$.
\end{enumerate}

A message $m$ \emph{causally precedes} another message $m'$, written $m \rightsquigarrow m'$, if the event in which $m$ is sent happens-before the event in which $m'$ is sent.
A pair of messages $m$ and $m'$ is \emph{concurrent}, denoted $m||m'$, if neither $m$ causally precedes $m'$ nor $m'$ causally precedes $m$.

\subparagraph{Leader oracle.}
Each replica is equipped with a local \emph{eventual leader oracle} $\Omega$~\cite{chandra1996weakest,cachin2011introduction}, which it queries at every algorithm step to receive a replica $\id$.
In any \emph{infinite} execution, the outputs produced by $\Omega$ satisfy:

\begin{itemize}
    \item (Eventual Accuracy) There is a step after which every query from a correct replica outputs the $\id$ of a correct replica.
    \item (Eventual Agreement) There is a step after which every query from all correct replicas outputs the same replica $\id$.
\end{itemize}

\subsection{Communication Layer}
\label{subsec:communication}
We restrict our attention to a class of protocols combining two components: a communication component, where each replica builds an ever-growing directed acyclic graph~\cite{gutin2009digraphs} (DAG) induced by the causal relations among messages, and a local computation component that interprets the DAG to produce outputs.
A message is a tuple $(\data,\id,\leader)$, where $\data$ is some message content, $\id$ is the identifier of the replica sending the message, and $\leader$ is the output of $\Omega$ at the event in which the message is sent.
For the communication component, every replica $r_i$ periodically (and forever) issues a new message $m$, in which case $r_i$ sends $m$ to every other replica $r_j$.
Upon receiving $m$, $r_j$ updates its local DAG by adding the new message as a \emph{vertex}.
Let $G_i$ denote the DAG at replica $r_i$ and let $\past_i(\bv)$ be the set of vertices $\bw$ such that there is a path from $\bw$ to $\bv$ in $G_i$. The construction satisfies the following properties in every infinite execution:

\begin{itemize}
    \item If a correct replica issues $\bv$, then every correct replica $r_j$ eventually adds $\bv$ to $G_j$.
    \item If a replica $r_i$ adds a vertex $\bv$ to $G_i$, then $\bv$ was previously issued by replica $\bv.\id$.
    \item For every vertex $\bv \in G_i$, $\bw \in \past_i(\bv)$ \textbf{iff} $\bw \rightsquigarrow \bv$.
    \item If a correct replica $r_i$ adds $\bv$ to $G_i$, then every correct replica $r_j$ eventually adds $\bv$ to $G_j$.
\end{itemize}

\subsection{Optimistic Replication with Commitment}
\label{subsec:specification}
Each replica maintains a local DAG reflecting the causality order of the exchanged messages.
Given the local DAG $G_i$ of a replica $r_i$, $r_i$ periodically turns its vertices into two sequences as outputs:
the optimistic ordering $\uncommittedS_i$ and the irrevocable prefix $\committedS_i$, defined as follows. 

\subparagraph{Optimistic ordering.} Let $\hist_i$ be the set of vertices in $G_i$. The optimistic sequence $\uncommittedS_i$ is a permutation of $\hist_i$. 
As new elements are added to the history, the order of some messages in the optimistic sequence may be revised, although this ordering must satisfy a convergence property and be consistent with the causal ordering of the messages.  
Formally, an optimistic replication protocol must satisfy:

\begin{itemize}
    \item (Strong Eventual-Consistency) For any pair of replicas $r_i$ and $r_j$, if $\hist_i = \hist_j$, then $\uncommittedS_i = \uncommittedS_j$.
    \item (Causal-Consistency) For every replica $r_i$ and all $m$ and $m'$ in $\hist_i$, if $m \rightsquigarrow m'$, then $m$ precedes $m'$ in $\uncommittedS_i$.
\end{itemize}

\subparagraph{Committed prefix.} On the other hand, the ordering of messages established in $\committedS_i$ should be \emph{irrevocable} and agreed upon by the replicas. Formally:

\begin{itemize}
    \item (Prefix Consistency) For any pair of replicas $r_i$ and $r_j$, $\committedS_i$ is a prefix of $\committedS_j$ or vice-versa.
    \item (Prefix Inclusion) For every replica $r_i$, $\committedS_i$ is a prefix of $\uncommittedS_i$.
    \item (Prefix Progress) For every correct replica, if $r_i$ adds $m$ to $\hist_i$, then $r_i$ eventually includes $m$ in $\committedS_i$.
\end{itemize}

In Section~\ref{sec:DAGframe}, we describe a method that uses properties of the induced graph to locally compute sequences satisfying the above properties.

\section{Deterministic Commitment}
\label{sec:DAGframe}
This section presents a deterministic commitment mechanism that derives a total order directly from the DAG.
The key idea is to leverage the information provided by the eventual leader oracle $\Omega$, which is embedded in every vertex. During periods in which $\Omega$ stabilizes, replicas can identify portions of the DAG that have been observed by a majority of replicas and use them to deterministically extend a common prefix.

Commitment proceeds in two steps. First, replicas apply a DAG predicate to identify a consistent sequence of special vertices, called \emph{pre-committed} vertices. They then use these vertices as reference points to order all vertices in their causal past according to a topological sorting $f$.\footnote{A topological sorting $f$ is a function that totally orders the vertices of a DAG, while preserving the partial order imposed by the edges. The reader may refer to~\cite{waitfree2025} for a construction of such a function.}

\begin{figure}
    \centering
    \begin{tikzpicture}[
            >=stealth,
            vertex/.style={inner sep=2pt},
            vote label/.style={
                font=\scriptsize,
                shape=circle,
                draw=black,
                line width=0.5pt,
                minimum size=9pt,
                inner sep=0.3pt,
                text=black
            },
            leader label/.style={
                font=\scriptsize,
                shape=rectangle,
                draw=black,
                line width=0.75pt,
                minimum size=9pt,
                inner sep=0.3pt,
                text=black,
                draw=black
            },
            hb/.style={->, dashed, line width=1pt},
            po/.style={->},
            lin/.style={->},
            causal region/.style={
                draw=none,
                fill=causalRegionColor!16,
                fill opacity=0.75,
                rounded corners=10pt,
                line join=round,
                blur shadow={
                    shadow blur radius=12pt,
                    shadow blur steps=24,
                    shadow xshift=0pt,
                    shadow yshift=0pt,
                    shadow opacity=12
                }
            },
            support halo/.style={
                shading=support radial,
                draw=none
            },
            support border/.style={
                draw=leaderSupportColor,
                fill=none,
                line width=0.8pt,
                rounded corners=8pt,
                line cap=round,
                line join=round
            },
            next event/.style={vertex},
            previous event/.style={vertex},
            leader event/.style={vertex},
            previous hatch/.style={
                pattern={Lines[angle=-45,distance=3pt,line width=0.6pt]},
                pattern color=previousColor!52
            },
            next hatch/.style={
                pattern={Lines[angle=45,distance=3pt,line width=0.6pt]},
                pattern color=nextColor!52
            },
        ]
        \begin{scope}[x=1cm]
        \def\causalregionpath{%
            (1.5,3.6)
            -- (6.5,3.6)
            -- (6.5,3.0)
            -- (5.5,2)
            -- (6.8,1.2)
            -- (6.5,-0.4)
            -- (-0.6,-0.4)
            -- (-0.6,0.5)
            -- (1,1)
            -- (2.5,2)
            -- (1.25,3)
            -- cycle
        }
        \def\supportregionpath{%
            (2.5,2.7)
            -- (5.5,2.7)
            -- (5.25,2)
            -- (6.5,1.35)
            -- (6.5,-0.25)
            -- (-0.5,-0.25)
            -- (-0.5,0.4)
            -- (1.08,0.95)
            -- (2.55,1.95)
            -- cycle
        }
        \path[causal region] \causalregionpath;
        \begin{scope}
            \clip[rounded corners=10pt] \causalregionpath;
            \clip[rounded corners=8pt] \supportregionpath;
            \shade[support halo] (4,1.2) circle[radius=4.8cm];
        \end{scope}
        \begin{scope}
            \clip[rounded corners=10pt] \causalregionpath;
            \path[previous hatch]
                plot[smooth, tension=0.48] coordinates {
                    (2.2,3) (3.4,2) (2.5,1) (1,0)
                }
                -- (-2,-1) -- (-2,4.5) -- (2.70,4.5) -- cycle;
            \path[next hatch]
                plot[smooth, tension=0.48] coordinates {
                    (5.5,3) (4.4,2) (5.7,1) (5.5,-0.4)
                }
                -- (9,-1) -- (9,4.5) -- (5.30,4.5) -- cycle;
        \end{scope}
        \begin{scope}
            \clip[rounded corners=10pt] \causalregionpath;
            \path[support border] \supportregionpath;
        \end{scope}

        \node[vertex] (a1) at (0,3) {$\mathbf{v}_{a:1}$};
        \node[previous event] (a2) at (2,3) {$\mathbf{v}_{a:2}$};
        \node[vertex] (a3) at (4,3) {$\mathbf{v}_{a:3}$};
        \node[next event] (a4) at (6,3) {$\mathbf{v}_{a:4}$};
        \node[vertex] (a5) at (8,3) {$\mathbf{v}_{a:5}$};

        \node[vertex] (b1) at (1,2) {$\mathbf{v}_{b:1}$};
        \node[previous event] (b2) at (3,2) {$\mathbf{v}_{b:2}$};
        \node[next event] (b3) at (5,2) {$\mathbf{v}_{b:3}$};
        \node[vertex] (b4) at (7,2) {$\mathbf{v}_{b:4}$};

        \node[vertex] (c1) at (0,1) {$\mathbf{v}_{c:1}$};
        \node[previous event] (c2) at (2,1) {$\mathbf{v}_{c:2}$};

        \node[leader event] (c3) at (4,1) {$\bv_{c:3}$};
        \node[next event] (c4) at (6,1) {$\mathbf{v}_{c:4}$};
        \node[vertex] (c5) at (8,1) {$\mathbf{v}_{c:5}$};

        \node[previous event] (d1) at (0,0) {$\mathbf{v}_{d:1}$};
        \node[vertex] (d2) at (2,0) {$\mathbf{v}_{d:2}$};
        \node[vertex] (d3) at (4,0) {$\mathbf{v}_{d:3}$};
        \node[next event] (d4) at (6,0) {$\mathbf{v}_{d:4}$};
        \node[vertex] (d5) at (8,0) {$\mathbf{v}_{d:5}$};

        \draw[po] (a1) -- (a2);
        \draw[po] (a2) -- (a3);
        \draw[po] (a3) -- (a4);
        \draw[po] (a4) -- (a5);

        \draw[po] (b1) -- (b2);
        \draw[po] (b2) -- (b3);
        \draw[po] (b3) -- (b4);

        \draw[po] (c1) -- (c2);
        \draw[po] (c2) -- (c3);
        \draw[po] (c3) -- (c4);
        \draw[po] (c4) -- (c5);

        \draw[po] (d1) -- (d2);
        \draw[po] (d2) -- (d3);
        \draw[po] (d3) -- (d4);
        \draw[po] (d4) -- (d5);

        \draw[hb] (a2) -- (b2);
        \draw[hb] (b2) -- (c3);
        \draw[hb] (c3) -- (b3);
        \draw[hb] (b3) -- (a4);
        \draw[hb] (d1) -- (c2);
        \draw[hb] (c3) -- (d4);
        \draw[hb] (c4) -- (d5);

        \node[vote label,above=1pt of a1] {$\votevalue{a}$};
        \node[vote label,above=1pt of a2] {$\votevalue{a}$};
        \node[vote label,above=1pt of a3] {$\votevalue{a}$};
        \node[vote label,above=1pt of a4] {$\votevalue{a}$};
        \node[vote label,above=1pt of a5] {$\votevalue{a}$};

        \node[vote label,above=1pt of b1] {$\votevalue{a}$};
        \node[vote label,above=1pt of b2] {$\votevalue{c}$};
        \node[vote label,above=1pt of b3] {$\votevalue{c}$};
        \node[vote label,above=1pt of b4] {$\votevalue{c}$};

        \node[vote label,above=1pt of c1] {$\votevalue{c}$};
        \node[vote label,above=1pt of c2] {$\votevalue{c}$};
        \node[leader label,above=1pt of c3] {$\votevalue{c}$};
        \node[vote label,above=1pt of c4] {$\votevalue{c}$};
        \node[vote label,above=1pt of c5] {$\votevalue{c}$};

        \node[vote label,above=1pt of d1] {$\votevalue{c}$};
        \node[vote label,above=1pt of d2] {$\votevalue{c}$};
        \node[vote label,above=1pt of d3] {$\votevalue{c}$};
        \node[vote label,above=1pt of d4] {$\votevalue{c}$};
        \node[vote label,above=1pt of d5] {$\votevalue{b}$};

        \node[font=\bfseries\footnotesize,anchor=south] at (4,1.55)
            {Leader};

        \end{scope}
        \replicarows
    \end{tikzpicture}

    \par\medskip
    \causallegend

    \caption{Local DAG of a replica $r_i$.
    The DAG represents $r_i$'s view at a fixed point in time, which might be incomplete.
    In this particular case, $\textbf{v}_{c:3}$ is a leader vertex.
    The vertex gathers support within its causal region from all vertices issued by replicas \textit{b}, \textit{c}, and \textit{d}, thus obtaining support from a majority of replicas.}
    \label{fig:dags1}
\end{figure}
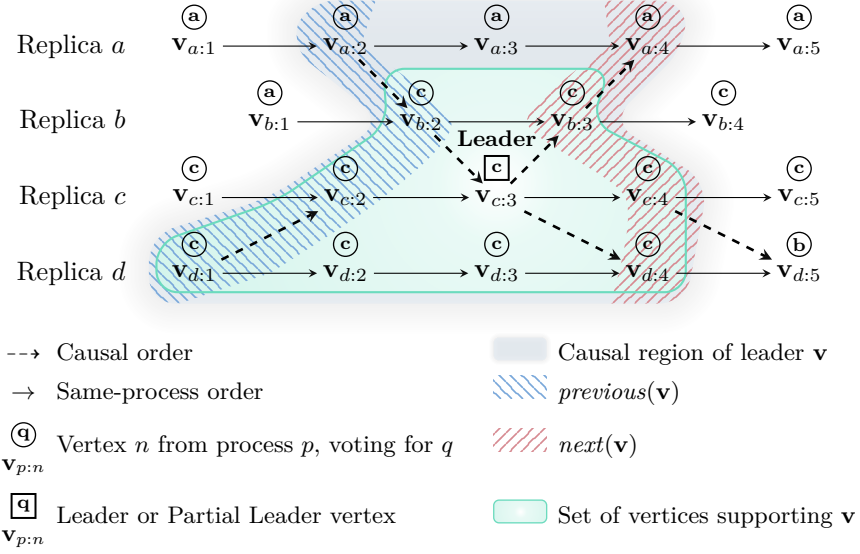

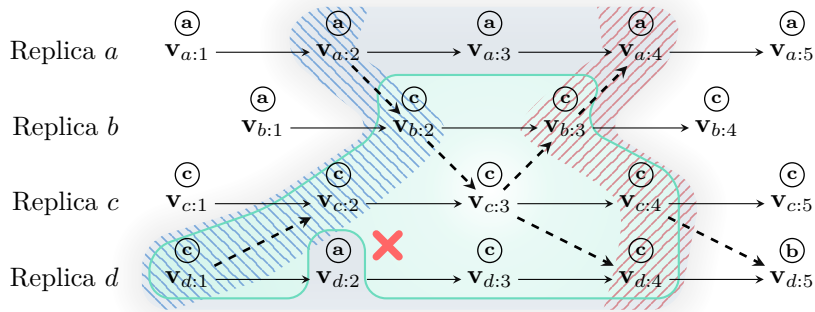
\begin{figure}
    \centering
    \begin{tikzpicture}[
            >=stealth,
            vertex/.style={inner sep=2pt},
            vote label/.style={
                font=\scriptsize,
                shape=circle,
                draw=black,
                line width=0.5pt,
                minimum size=9pt,
                inner sep=0.3pt,
                text=black
            },
            leader label/.style={
                font=\scriptsize,
                shape=rectangle,
                draw=black,
                line width=0.75pt,
                minimum size=9pt,
                inner sep=0.3pt,
                text=black,
                draw=black
            },
            hb/.style={->, dashed, line width=1pt},
            po/.style={->},
            lin/.style={->},
            causal region/.style={
                draw=none,
                fill=causalRegionColor!16,
                fill opacity=0.75,
                rounded corners=10pt,
                line join=round,
                blur shadow={
                    shadow blur radius=12pt,
                    shadow blur steps=24,
                    shadow xshift=0pt,
                    shadow yshift=0pt,
                    shadow opacity=12
                }
            },
            support halo/.style={
                shading=support radial,
                draw=none
            },
            support border/.style={
                draw=leaderSupportColor,
                fill=none,
                line width=0.8pt,
                rounded corners=8pt,
                line cap=round,
                line join=round
            },
            next event/.style={vertex},
            previous event/.style={vertex},
            leader event/.style={vertex},
            previous hatch/.style={
                pattern={Lines[angle=-45,distance=3pt,line width=0.6pt]},
                pattern color=previousColor!52
            },
            next hatch/.style={
                pattern={Lines[angle=45,distance=3pt,line width=0.6pt]},
                pattern color=nextColor!52
            },
            BarreStyle/.style={line width=1 mm,line cap=round,color=red!60},
        ]
        \begin{scope}[x=1cm]
        \def\causalregionpath{%
            (1.5,3.6)
            -- (6.5,3.6)
            -- (6.5,3.0)
            -- (5.5,2)
            -- (6.8,1.2)
            -- (6.5,-0.4)
            -- (-0.6,-0.4)
            -- (-0.6,0.5)
            -- (1,1)
            -- (2.5,2)
            -- (1.25,3)
            -- cycle
        }
        \def\supportregionpath{%
            (2.5,2.7)
            -- (5.5,2.7)
            -- (5.25,2)
            -- (6.5,1.35)
            -- (6.5,-0.25)
            -- (2.35, -0.25)
            -- (2.35, 0.65)
            -- (1.6, 0.65)
            -- (1.6, -0.25)
            -- (-0.5,-0.25)
            -- (-0.5,0.4)
            -- (1.08,0.95)
            -- (2.55,1.95)
            -- cycle
        }
        \path[causal region] \causalregionpath;
        \begin{scope}
            \clip[rounded corners=10pt] \causalregionpath;
            \clip[rounded corners=8pt] \supportregionpath;
            \shade[support halo] (4,1.2) circle[radius=4.8cm];
        \end{scope}
        \begin{scope}
            \clip[rounded corners=10pt] \causalregionpath;
            \path[previous hatch]
                plot[smooth, tension=0.48] coordinates {
                    (2.2,3) (3.4,2) (2.5,1) (1,0)
                }
                -- (-2,-1) -- (-2,4.5) -- (2.70,4.5) -- cycle;
            \path[next hatch]
                plot[smooth, tension=0.48] coordinates {
                    (5.5,3) (4.4,2) (5.7,1) (5.5,-0.4)
                }
                -- (9,-1) -- (9,4.5) -- (5.30,4.5) -- cycle;
        \end{scope}
        \begin{scope}
            \clip[rounded corners=10pt] \causalregionpath;
            \path[support border] \supportregionpath;
        \end{scope}

        \node[vertex] (a1) at (0,3) {$\mathbf{v}_{a:1}$};
        \node[previous event] (a2) at (2,3) {$\mathbf{v}_{a:2}$};
        \node[vertex] (a3) at (4,3) {$\mathbf{v}_{a:3}$};
        \node[next event] (a4) at (6,3) {$\mathbf{v}_{a:4}$};
        \node[vertex] (a5) at (8,3) {$\mathbf{v}_{a:5}$};

        \node[vertex] (b1) at (1,2) {$\mathbf{v}_{b:1}$};
        \node[previous event] (b2) at (3,2) {$\mathbf{v}_{b:2}$};
        \node[next event] (b3) at (5,2) {$\mathbf{v}_{b:3}$};
        \node[vertex] (b4) at (7,2) {$\mathbf{v}_{b:4}$};

        \node[vertex] (c1) at (0,1) {$\mathbf{v}_{c:1}$};
        \node[previous event] (c2) at (2,1) {$\mathbf{v}_{c:2}$};

        \node[leader event] (c3) at (4,1) {$\bv_{c:3}$};
        \node[next event] (c4) at (6,1) {$\mathbf{v}_{c:4}$};
        \node[vertex] (c5) at (8,1) {$\mathbf{v}_{c:5}$};

        \node[previous event] (d1) at (0,0) {$\mathbf{v}_{d:1}$};
        \node[vertex] (d2) at (2,0) {$\mathbf{v}_{d:2}$};
        \node[vertex] (d3) at (4,0) {$\mathbf{v}_{d:3}$};
        \node[next event] (d4) at (6,0) {$\mathbf{v}_{d:4}$};
        \node[vertex] (d5) at (8,0) {$\mathbf{v}_{d:5}$};

        \draw[po] (a1) -- (a2);
        \draw[po] (a2) -- (a3);
        \draw[po] (a3) -- (a4);
        \draw[po] (a4) -- (a5);

        \draw[po] (b1) -- (b2);
        \draw[po] (b2) -- (b3);
        \draw[po] (b3) -- (b4);

        \draw[po] (c1) -- (c2);
        \draw[po] (c2) -- (c3);
        \draw[po] (c3) -- (c4);
        \draw[po] (c4) -- (c5);

        \draw[po] (d1) -- (d2);
        \draw[po] (d2) -- (d3);
        \draw[po] (d3) -- (d4);
        \draw[po] (d4) -- (d5);

        \draw[hb] (a2) -- (b2);
        \draw[hb] (b2) -- (c3);
        \draw[hb] (c3) -- (b3);
        \draw[hb] (b3) -- (a4);
        \draw[hb] (d1) -- (c2);
        \draw[hb] (c3) -- (d4);
        \draw[hb] (c4) -- (d5);

        \node[vote label,above=1pt of a1] {$\votevalue{a}$};
        \node[vote label,above=1pt of a2] {$\votevalue{a}$};
        \node[vote label,above=1pt of a3] {$\votevalue{a}$};
        \node[vote label,above=1pt of a4] {$\votevalue{a}$};
        \node[vote label,above=1pt of a5] {$\votevalue{a}$};

        \node[vote label,above=1pt of b1] {$\votevalue{a}$};
        \node[vote label,above=1pt of b2] {$\votevalue{c}$};
        \node[vote label,above=1pt of b3] {$\votevalue{c}$};
        \node[vote label,above=1pt of b4] {$\votevalue{c}$};

        \node[vote label,above=1pt of c1] {$\votevalue{c}$};
        \node[vote label,above=1pt of c2] {$\votevalue{c}$};
        \node[vote label,above=1pt of c3] {$\votevalue{c}$};
        \node[vote label,above=1pt of c4] {$\votevalue{c}$};
        \node[vote label,above=1pt of c5] {$\votevalue{c}$};

        \node[vote label,above=1pt of d1] {$\votevalue{c}$};
        \node[vote label,above=1pt of d2] {$\votevalue{a}$};
        \node[vote label,above=1pt of d3] {$\votevalue{c}$};
        \node[vote label,above=1pt of d4] {$\votevalue{c}$};
        \node[vote label,above=1pt of d5] {$\votevalue{b}$};

        \draw[BarreStyle] (2.5, 0.3) to (2.8, 0.6);
        \draw[BarreStyle] (2.8, 0.3) to (2.5, 0.6);

        \end{scope}
        \replicarows
        \end{tikzpicture}
    \caption{
        Local DAG without a leader vertex.
        The only difference from Figure~\ref{fig:dags1} is the leader output of $v_{d:2}$.
        The vertex $\textbf{v}_{c:3}$ does not gather sufficient support within its causal region, since $v_{d:2}$ supports replica \textit{a} rather than replica \textit{c}.
    }
    \label{fig:dags2}
\end{figure}

\begin{figure}
    \centering
    \begin{tikzpicture}[
            >=stealth,
            vertex/.style={inner sep=2pt},
            vote label/.style={
                font=\scriptsize,
                shape=circle,
                draw=black,
                line width=0.5pt,
                minimum size=9pt,
                inner sep=0.3pt,
                text=black
            },
            leader label/.style={
                font=\scriptsize,
                shape=rectangle,
                draw=black,
                line width=0.75pt,
                minimum size=9pt,
                inner sep=0.3pt,
                text=black,
                draw=black
            },
            hb/.style={->, dashed, line width=1pt},
            po/.style={->},
            lin/.style={->},
            causal region/.style={
                draw=none,
                fill=causalRegionColor!16,
                fill opacity=0.75,
                rounded corners=10pt,
                line join=round,
                blur shadow={
                    shadow blur radius=12pt,
                    shadow blur steps=24,
                    shadow xshift=0pt,
                    shadow yshift=0pt,
                    shadow opacity=12
                }
            },
            support halo/.style={
                shading=support radial,
                draw=none
            },
            support border/.style={
                draw=leaderSupportColor,
                fill=none,
                line width=0.8pt,
                rounded corners=8pt,
                line cap=round,
                line join=round
            },
            next event/.style={vertex},
            previous event/.style={vertex},
            leader event/.style={vertex},
            previous hatch/.style={
                pattern={Lines[angle=-45,distance=3pt,line width=0.6pt]},
                pattern color=previousColor!52
            },
            next hatch/.style={
                pattern={Lines[angle=45,distance=3pt,line width=0.6pt]},
                pattern color=nextColor!52
            },
        ]
        \begin{scope}[x=0.81632653cm]

        \def\partialregionpath{%
            (-0.6,3.6) -- (0.6,3.6) -- (2.2,2.7)
            -- (3.8,2.7) -- (3.8,0.6) -- (-0.6,0.6) -- cycle
        }
        \def\causalregionpath{%
            (6,2.7) -- (10.4,2.7) -- (10.4,-0.4)
            -- (4.4,-0.4) -- (4.4,0.6) -- (6,1.6) -- cycle
        }
        \foreach \region in {\partialregionpath,\causalregionpath} {
            \path[causal region] \region;
        }
        \begin{scope}
            \clip[rounded corners=10pt] \partialregionpath;
            \shade[support halo] (1.6,2.2) circle[radius=4.8cm];
            \path[previous hatch]
                plot[smooth,tension=0.48] coordinates {
                    (0.5,3.6) (0.5,3) (0.5,2) (0.5,0.6)
                } -- (-1,0.6) -- (-1,3.6) -- cycle;
            \path[next hatch]
                (2.65,2.7) -- (2.65,0.6) -- (4,0.6) -- (4,2.7) -- cycle;
        \end{scope}
        \path[support border] \partialregionpath;
        \begin{scope}
            \clip[rounded corners=10pt] \causalregionpath;
            \shade[support halo] (8.2,1.2) circle[radius=4.8cm];
            \path[previous hatch]
                plot[smooth,tension=0.48] coordinates {
                    (7.1,2.7) (7.1,2) (7.1,1) (5.5,0) (5.5,-0.4)
                } -- (3.8,-0.4) -- (3.8,2.7) -- cycle;
            \path[next hatch]
                (9.25,2.7) -- (9.25,-0.4) -- (10.8,-0.4) -- (10.8,2.7) -- cycle;
        \end{scope}
        \path[support border] \causalregionpath;

        \node[previous event] (a1) at (0,3) {$\mathbf{v}_{a:1}$};

        \node[previous event] (b1) at (0,2) {$\mathbf{v}_{b:1}$};
        \node[leader event] (b2) at (1.6,2) {$\bv_{b:2}$};
        \node[next event] (b3) at (3.2,2) {$\mathbf{v}_{b:3}$};
        \node[previous event] (b4) at (6.6,2) {$\mathbf{v}_{b:4}$};
        \node[next event] (b5) at (9.8,2) {$\mathbf{v}_{b:5}$};

        \node[previous event] (c1) at (0,1) {$\mathbf{v}_{c:1}$};
        \node[next event] (c2) at (3.2,1) {$\mathbf{v}_{c:2}$};
        \node[previous event] (c3) at (6.6,1) {$\mathbf{v}_{c:3}$};
        \node[leader event] (c4) at (8.2,1) {$\bv_{c:4}$};
        \node[next event] (c5) at (9.8,1) {$\mathbf{v}_{c:5}$};

        \node[previous event] (d1) at (5,0) {$\mathbf{v}_{d:1}$};
        \node[next event] (d2) at (9.8,0) {$\mathbf{v}_{d:2}$};

        \draw[po] (b1) -- (b2);
        \draw[po] (b2) -- (b3);
        \draw[po] (b3) -- (b4);
        \draw[po] (b4) -- (b5);
        \draw[po] (c1) -- (c2);
        \draw[po] (c2) -- (c3);
        \draw[po] (c3) -- (c4);
        \draw[po] (c4) -- (c5);
        \draw[po] (d1) -- (d2);

        \draw[hb] (a1) -- (b2);
        \draw[hb] (c1) -- (b2);
        \draw[hb] (b2) -- (c2);
        \draw[hb] (b4) -- (c4);
        \draw[hb] (c4) -- (b5);
        \draw[hb] (c4) -- (d2);
        \draw[hb] (d1) -- (c3);

        \foreach \v in {a1,b1,b3,c1,c2} {
            \node[vote label,above=1pt of \v] {$\votevalue{b}$};
        }
        \foreach \v in {b4,b5,c3,c5,d1,d2} {
            \node[vote label,above=1pt of \v] {$\votevalue{c}$};
        }
        \node[leader label,above=1pt of b2] {$\votevalue{b}$};
        \node[leader label,above=1pt of c4] {$\votevalue{c}$};
        \node[font=\bfseries\footnotesize,anchor=south] at (1.9,2.7)
            {Partial leader};
        \node[font=\bfseries\footnotesize,anchor=south] at (8.2,1.55)
            {Leader};
        \end{scope}
        \replicarows
        \end{tikzpicture}
    \caption{
        Local DAG with leader vertex $v_{c:4}$, where $v_{b:2}$ is a partial leader in the past of $v_{c:4}$.
        The vertex $\textbf{v}_{b:2}$ can still gather support from a majority of replicas, for instance through an unseen vertex from replica \textit{a}, so it is a partial leader from $v_{c:4}$'s perspective (when looking in the past of $v_{c:4}$).
    }
    \label{fig:dags3}
\end{figure}
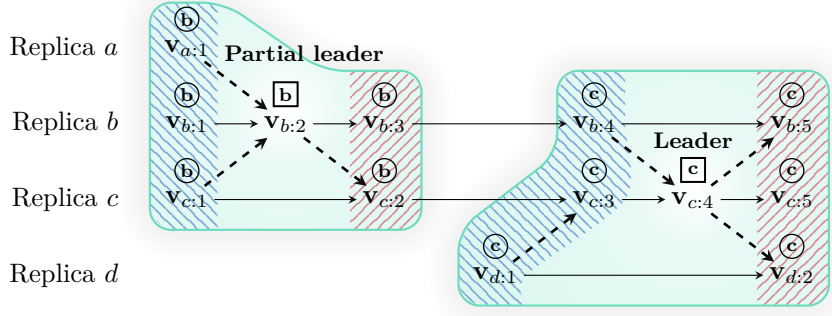

Intuitively, a vertex becomes pre-committed if it gathers support from ``enough'' vertices in its past and future.\footnote{A vertex $\bv$ is in the future of a vertex $\bw$ if $\bw \in \past(\bv)$.}
Each $\leader$ identifier (output by $\Omega$) associated with a vertex $\bv$ counts as a vote for a vertex $\bw$ issued by $r_{\leader}$.
If a majority of replicas vote for $\bw.\id$ (the issuer of $\bw$) in the \emph{causal region} of $\bw$, then $\bw$ becomes a \emph{leader vertex} and is pre-committed.
As illustrated in Figures~\ref{fig:dags1} to~\ref{fig:dags3}, the causal region of $\bw$ is composed of its immediate past\footnote{The precise definitions are given in Section~\ref{subsec:dagPred}.} $\previous(\bw)$ (the latest vertices from each replica seen by $\bw$), its immediate future $\nnext(\bw)$ (the earliest vertices from each replica to see $\bw$), and vertices ``in between''.

A vertex in the past of a pre-committed vertex is also considered pre-committed if it could eventually gather sufficient support to become a leader.
Such a vertex is referred to as a \emph{partial leader}.
This condition ensures that future pre-committed vertices cannot miss vertices that were previously committed, as shown in Section~\ref{subsec:dagPred}.

\subsection{DAG Predicate}
\label{subsec:dagPred}
We now describe the construction of a function $\cV$ that extracts a sequence of pre-committed vertices $\cV(G_i)$ (or $\cV_i$ for short) from a local DAG $G_i$.
For all replicas $r_i$ and $r_j$ and all execution steps, the construction should satisfy:

\begin{enumerate}
    \item $\cV_i \preceq \cV_j$ or vice-versa.
    \item If $G_i$ is a subgraph of $G_j$, then $\cV_i \preceq \cV_j$.
    \item Every vertex in $G_i$ eventually appears in the past of some vertex appearing in $\cV_i$.
\end{enumerate}

In the definitions and statements below, we consider a fixed point in time in an arbitrary run. 
We omit subscripts when there is no ambiguity, or when concepts apply to all graphs.

\subparagraph{Preliminaries.}
For $\bv \in G$, $\nnext(\bv)$ denotes the set of vertices $\bw \in G$ such that $\bw$ is the \emph{first} vertex from $\bw.\id$ satisfying $\bv \rightsquigarrow \bw$.
Analogously, $\previous(\bv)$ denotes the set of vertices $\bz \in G$ such that $\bz$ is the \emph{last} vertex from $\bz.\id$ satisfying $\bz \rightsquigarrow \bv$.

Let $W$ be a set of vertices and let $\rep(W)$ be the set of replicas with at least one issued vertex appearing in $W$.
The set $\votes(r_i,W)$ is comprised of replicas $r_j \in \rep(W)$ such that every vertex from $r_j$ in $W$ has $r_i$ as the leader.
A \emph{projection} of $W$ over a set of replicas $P$, written $W|_P$, is the set of vertices $\bv \in W$ such that $\bv.\id \in P$.

Now, we give the definition of the set of vertices $\current(\bv)$.
A vertex $\bw$ belongs to $\current(\bv)$ \textbf{iff} $\bw.\id \in \rep(\previous(\bv))$ (there is a vertex from the issuer of $\bw$ in $\previous(\bv)$) \emph{and} one of the two following conditions is satisfied:

\begin{enumerate}
    \item $\bw \in \nnext(\bv)$.
    \item $\not\exists \bz \in \nnext(\bv)$ such that $\bz.\id = \bw.\id$, and $\bw$ is the last vertex issued by $\bw.\id$ in $G$.
\end{enumerate}

Let $\bv,\bw \in G$ be two vertices issued by the \emph{same replica} $r$.
Then, $\trace(\bv, \bw)$ is the set comprised of all vertices from $\bv$ to $\bw$ (inclusive) that were issued by $r$.
If $\bv.\id \neq \bw.\id$ or $\bw \rightsquigarrow \bv$ ($\bw$ was issued before $\bv$), we define $\trace(\bv, \bw) \triangleq \emptyset$.
We say that $\trace(\bv, \bw)$ \emph{precedes} $\trace(\bv',\bw')$ if both are non-empty and $\bw \rightsquigarrow \bv'$.

The trace-closure of $W$, written $\cT(W)$, is the smallest set containing all possible traces for pairs of vertices in $W$, and is formally defined as: $\cT(W) \triangleq \bigcup_{\bv,\bw \in W}\trace(\bv,\bw)$.

We define the causal region of $\bv$ as $\causalRegion(\bv) \triangleq \cT(\previous(\bv) \cup \current(\bv))$.
Consider the first vertex $\bw$ and last vertex $\bz$ issued by a replica $r$ in $\causalRegion(\bv)$.
Observe that $\causalRegion(\bv)|_{\{r\}} = \trace(\bw, \bz)$, i.e., $\causalRegion(\bv)|_{\{r\}}$ is the trace of $r$ in the causal region of $\bv$.
In the following, a \emph{quorum} is a set with at least a majority of replicas.

\begin{definition}[Leader]
\label{def:leader}
Let $P$ be a set of replicas and $\bv$ a vertex.
We say that $\bv$ is a \emph{$P$-leader} if all the following requirements are satisfied:

\begin{enumerate}
    \item $P \subseteq \rep(\previous(\bv))$. There is a vertex in $\previous(\bv)$ for every replica belonging to $P$.
    \item $P \subseteq \rep(\nnext(\bv))$. There is a vertex in $\nnext(\bv)$ for every replica belonging to $P$.
    \item $P \subseteq \votes(\bv.\id, \causalRegion(\bv))$. All vertices in the causal region of $\bv$ (that were issued by replicas belonging to $P$) vote for $\bv.\id$.
\end{enumerate}

We say that $\bv$ is a \emph{leader} vertex if there exists a quorum $Q$ such that $\bv$ is a $Q$-leader.
\end{definition}

\begin{lemma}
\label{lm:parallelLeader}
Let $\bv$ be a leader vertex in $G_i$ and $\bw$ be a leader vertex in $G_j$ such that $\bv \neq \bw$,
then either $\bv \rightsquigarrow \bw$ or $\bw \rightsquigarrow \bv$.
\begin{proof}
If $\bv.\id = \bw.\id$, then the result is immediate.
Otherwise, let $Q$ be a quorum for which $\bv$ is a $Q$-leader in $G_i$, and let $Q'$ be a quorum for which $\bw$ is a $Q'$-leader in $G_j$.
Take $r \in Q \cap Q'$.
By definition of a leader, there exist $\bz_{\prev} \in \previous(\bv)$ and $\bz_{\nnext} \in \nnext(\bv)$ issued by $r$ in $G_i$,
but also $\bz'_{\prev} \in \previous(\bw)$ and $\bz'_{\nnext} \in \nnext(\bw)$ issued by $r$ in $G_j$.
Observe that $\bz_{\prev} \rightsquigarrow \bv \rightsquigarrow \bz_{\nnext}$ and $\bz'_{\prev} \rightsquigarrow \bw \rightsquigarrow \bz'_{\nnext}$.
Since all vertices in $\causalRegion(\bv)|_{\{r\}}$ vote for $\bv.\id$ and all vertices in $\causalRegion(\bw)|_{\{r\}}$ vote for $\bw.\id$,
where $\bv.\id \neq \bw.\id$, the two traces are disjoint.
Because those vertices are issued by the same replica $r$, either $\bz_{\nnext} \rightsquigarrow \bz'_{\prev}$ or $\bz'_{\nnext} \rightsquigarrow \bz_{\prev}$, and the result follows from the previous observation.
\end{proof}
\end{lemma}

Lemma~\ref{lm:parallelLeader} implies that if a vertex $\bv$ is a leader in graph $G_i$, then there is no vertex $\bw$ parallel to $\bv$ that is also a leader in any graph $G_j$.

A vertex $\bv \in G$ may contain an \emph{incomplete causal region} for a quorum $Q$ voting for $\bv.\id$,
so that $\bv$ can potentially become a leader after adding vertices to $G$.
We capture this notion with the definition of \emph{partial leader}.

\begin{definition}[Partial Leader]
\label{def:partialLeader}
A vertex $\bv \in G$ is a \emph{partial leader} if there exists a quorum $Q$ such that both conditions hold:

\begin{enumerate}
    \item $Q \subseteq \rep(\previous(\bv))$. There is a vertex in $\previous(\bv)$ for every replica belonging to $Q$.
    \item $Q \subseteq \votes(\bv.\id, \causalRegion(\bv))$. All vertices in the causal region of $\bv$ (that were issued by replicas belonging to $Q$) vote for $\bv.\id$.
\end{enumerate}

If the conditions above hold, we say that $\bv$ is a partial leader for $Q$.
Note that a $Q$-leader vertex is also a partial leader for $Q$.
\end{definition}

\begin{lemma}
\label{lm:partialLeader}
If $\bv$ and $\bw$ are vertices such that $\bv \neq \bw$ and both are partial leaders in the same graph $G$, then either $\bv \rightsquigarrow \bw$ or $\bw \rightsquigarrow \bv$.

\begin{proof}
The result trivially holds in the case with $\bv.\id = \bw.\id$.
Otherwise, let $Q$ and $Q'$ be quorums for which $\bv$ and $\bw$ are partial leaders respectively.
Take $r \in Q \cap Q'$.
By definition, there exist $\bz_{\prev} \in \previous(\bv)$ and $\bz'_{\prev} \in \previous(\bw)$ issued by $r$.
Also, let $\bz_{\llast}$ and $\bz'_{\llast}$ be the \emph{last} vertices issued by $r$ in $\causalRegion(\bv)$ and $\causalRegion(\bw)$ respectively.
Since every vertex in $\trace(\bz_{\prev}, \bz_{\llast})$ (recall that $\causalRegion(\bv)|_{\{r\}} = \trace(\bz_{\prev}, \bz_{\llast})$) votes for $\bv.\id$ and every vertex in $\trace(\bz'_{\prev}, \bz'_{\llast})$ votes for $\bw.\id$,
where $\bv.\id \neq \bw.\id$, the two traces are disjoint and one must precede the other.
Assume w.l.o.g. that $\trace(\bz_{\prev}, \bz_{\llast})$ precedes $\trace(\bz'_{\prev}, \bz'_{\llast})$ and suppose that $\bw || \bv$.
If $\bz_{\llast} \in \nnext(\bv)$, then $\bv \rightsquigarrow \bz_{\llast} \rightsquigarrow \bz'_{\prev} \rightsquigarrow \bw$,
which contradicts the assumption that $\bw || \bv$.
On the other hand, if $\bz_{\llast} \not\in \nnext(\bv)$, then by the definitions of $\current(\bv)$ and trace closure, all vertices issued by $r$ after $\bz_{\prev}$ (including those within $\trace(\bz'_{\prev}, \bz'_{\llast})$) should belong to $\trace(\bz_{\prev}, \bz_{\llast})$.
This contradicts the fact that $\trace(\bz_{\prev}, \bz_{\llast})$ and $\trace(\bz'_{\prev}, \bz'_{\llast})$ are disjoint.
\end{proof}
\end{lemma}

From now on, we will write $\causalRegion_i(\bv)$ (with the subscript $i$) to denote the causal region of $\bv$ according to a specific graph $G_i$.

\begin{lemma}
\label{lm:alwaysLeader}
Let $\bv$ be a leader in $G_i$. Then $\bv$ is a partial leader in any $G_j$ such that $\bv \in G_j$.
\begin{proof}
Let $Q$ be a quorum for which $\bv$ is a $Q$-leader in $G_i$.
Because $Q \subseteq \rep(\nnext(\bv))$, the set $\causalRegion_i(\bv)|_{Q}$ contains the \emph{complete} traces of every replica belonging to $Q$, with every trace ending at a vertex in $\nnext(\bv)$ (observe that all traces start from a vertex in $\previous(\bv)$ regardless of the graph).
Moreover, all vertices in $\causalRegion_i(\bv)|_{Q}$ vote for $\bv.\id$.
Take any replica $r \in Q$.
For any other graph $G_j$ containing $\bv$, the trace of $r$ in $\causalRegion_j(\bv)$ is either incomplete or it ends at $\nnext(\bv)|_{\{r\}}$. Therefore it is a subset of the trace in $\causalRegion_i(\bv)$.
Also, for all replicas in $Q$, their traces vote for $\bv.\id$. Thus $\bv$ satisfies the conditions to be a partial leader.
\end{proof}
\end{lemma}

\begin{proposition}
\label{prop:subgraph}
Consider two graphs $G_i$ and $G_j$ at replicas $r_i$ and $r_j$ respectively.
Let $G_i \subseteq G_j$.
For every vertex $\bv$ appearing in both graphs: $\causalRegion_i(\bv) \subseteq \causalRegion_j(\bv)$.
\begin{proof}
Because $G_i \subseteq G_j$, $G_j$ contains every vertex (and edges) belonging to $G_i$.
Moreover, from the properties of the communication component, $\past(\bw)$ is the same for every vertex $\bw$ appearing in both graphs.
In particular, $\past(\bv)$ is the same in $G_i$ and $G_j$.
Consider any replica $r$ with a vertex in $\causalRegion_i(\bv)$.
The trace of $r$ in $\causalRegion_i(\bv)$ starts from $\previous(\bv)|_{\{r\}}$ and ends at a vertex $\bz$.
Let $\bz'$ be the last vertex from $r$ in $\causalRegion_j(\bv)$.
It follows that either $\bz = \bz'$ or $\bz \rightsquigarrow \bz'$, thus $\causalRegion_i(\bv)|_{\{r\}} \subseteq \causalRegion_j(\bv)|_{\{r\}}$.
And since this holds for every such replica $r$, $\causalRegion_i(\bv) \subseteq \causalRegion_j(\bv)$.
\end{proof}
\end{proposition}

\begin{lemma}
\label{lm:retroleader}
Consider two graphs $G_i$ and $G_j$ at replicas $r_i$ and $r_j$ respectively, such that $G_i \subseteq G_j$.
Let $\bv$ be a vertex that appears in both $G_i$ and $G_j$.
If $\bv$ is a partial leader in $G_j$, then it is also a partial leader in $G_i$.
\begin{proof}
Let $Q$ be a quorum for which $\bv$ is a partial leader in $G_j$.
Observe that whenever a graph contains $\bv$, it holds that $Q \subseteq \rep(\previous(\bv))$.
Since $G_i$ is a subgraph of $G_j$, $\causalRegion_i(v)|_{Q} \subseteq \causalRegion_j(v)|_{Q}$ (by Proposition~\ref{prop:subgraph}).
Moreover, because $\bv$ is a partial leader in $G_j$, all vertices in $\causalRegion_j(v)|_{Q}$ vote for $\bv.\id$.
Therefore, all vertices in $\causalRegion_i(v)|_{Q}$ also vote for $\bv.\id$, which implies that $\bv$ is a partial leader in $G_i$.
\end{proof}
\end{lemma}

Let $V$ be a set of vertices, the \emph{projection} of $G$ onto $V$, denoted $G|_{V}$, is the \emph{induced subgraph}\footnote{A graph $G$ is an \emph{induced subgraph} of $H$ if $G \subseteq H$ and $G$ contains all edges belonging to $H$ such that both endpoints are vertices of $G$~\cite{gutin2009digraphs}.} of $G$ containing only the vertices belonging to $V$.
We define $\graph(\bv) \triangleq G|_{\past(\bv) \cup \{\bv\}}$.\footnote{Note that $\graph_i(\bv)$ is the same for every graph $G_i$ containing $\bv$.}

\begin{definition}[Pre-Committed Set]
\label{def:comSet}
Let $\isLeader(\bv)$ be the predicate: $\bv$ is a leader vertex in $G$.
Also, let $\isPartial(\bv, \bw)$ be the predicate: $\bv$ is a partial leader in $\graph(\bw)$.
The set $\cC$ of pre-committed vertices of $G$ is built according to Algorithm~\ref{alg:preCommit}.
The algorithm computes $\cC$ by applying the following rules:

\begin{enumerate}
    \item Add every leader vertex $\bv$ to $\cC$.
    \item Recursively add $\bv$ to $\cC$ such that $\isPartial(\bv, \bw)$ holds for any $\bw \in \cC$.
\end{enumerate}

\begin{algorithm}
\begin{smartalgorithmic}[1]

    \Operation{\getPreCommitted}{$G$}
        \State $\oldC \gets \emptyset$
        \State $\newC \gets \{\bv \ | \ \isLeader(\bv)\}$ \label{ln:addLeaders}
        \While{$\oldC \neq \newC$}
            \State $\oldC \gets \newC$
            \State $\newC \gets \{\bv \ | \ \exists \bw \in \oldC: \isPartial(\bv, \bw)\}$ \label{ln:getPartials}
            \State $\newC \gets \newC \cup \oldC$ \label{ln:addPartials}
        \EndWhile
        \State \Return{$\newC$}
    \EndOperation

\end{smartalgorithmic}
\caption{Code for replica $r$ - Computation of pre-committed set.}
\label{alg:preCommit}
\end{algorithm}
\end{definition}

\begin{lemma}
\label{lm:pastLeader}
Suppose that $\isLeader(\bv)$ hold in $G_i$ for some vertex $\bv$ and $\isLeader(\bw)$ hold in $G_j$ for some vertex $\bw$. Then either $\isPartial(\bw,\bv)$ holds in $G_i$ or $\isPartial(\bv,\bw)$ holds in $G_j$.
\begin{proof}
Both $\bv$ and $\bw$ are leaders. Thus from Lemma~\ref{lm:parallelLeader}, either $\bv \rightsquigarrow \bw$ or $\bw \rightsquigarrow \bv$.
Assume w.l.o.g. that $\bw \rightsquigarrow \bv$. Then $\bw \in \past(\bv)$ and by Lemma~\ref{lm:alwaysLeader}, $\bw$ is also a partial leader in $\graph(\bv)$, meaning $\isPartial(\bw,\bv)$ holds in $G_i$.
\end{proof}
\end{lemma}

\begin{lemma}
\label{lm:totalC}
The set $\cC$ of pre-committed vertices of a graph is totally ordered by $\rightsquigarrow$.
\begin{proof}
Consider a particular call of $\getPreCommitted(G)$ from Algorithm~\ref{alg:preCommit}.
We split the call into rounds based on iterations of the while loop, where round $0$ is the execution of line~\ref{ln:addLeaders}.
Let $\newC^k$ be the set $\newC$ at the end of round $k$ (either lines~\ref{ln:addLeaders} or~\ref{ln:addPartials}).
We show by induction on the number of rounds that $\newC^k$ is totally ordered by $\rightsquigarrow$.

At the end of round $0$, $\newC^0$ is entirely comprised of leader vertices.
Thus, by Lemma~\ref{lm:parallelLeader}, $\newC^0$ is totally ordered by $\rightsquigarrow$.

Assume that by the end of round $n$, $\newC^n$ is totally ordered by $\rightsquigarrow$.
Suppose that there exist $\bv,\bw \in \newC^{n+1}$ such that $\bv || \bw$.
There are two cases to consider: either (1) (w.l.o.g.) $\bv$ was first added to $\newC$ in round $n+1$ and $\bw$ was added to $\newC$ in an older round,
or (2) both $\bv$ and $\bw$ were added to $\newC$ in round $n+1$.
We address each case in the following.

\subparagraph{Case (1).}
$\bv$ was added to $\newC$ in round $n+1$ and $\bw$ was added in a previous round.
Let $\bw^k$ be a vertex satisfying:

\begin{enumerate}
    \item $\bw^k || \bv$.
    \item $\bw^k$ was added to $\newC$ in round $k$ ($k < n+1$).
    \item $k$ is the \emph{smallest} round in which there exists a vertex $\bw'$, added to $\newC$ in round $k$, such that $\bw' || \bv$.
\end{enumerate}

Since $\bv$ was added to $\newC$ in round $n+1$, by construction of Algorithm~\ref{alg:preCommit}, there exists a vertex $\bv^n$ (first added in round $n$) for which $\isPartial(\bv,\bv^n)$ holds,
and $\bv^n$ was used in line~\ref{ln:getPartials} to add $\bv$ to $\newC$.
We say that $\bv$ was added to $\newC$ from the past of $\bv^n$.
It follows that $\bv \rightsquigarrow \bv^n$ and that $\bv$ is a partial leader in $\graph(\bv^n)$.

Because both $\bv^n$ and $\bw^k$ belong to $\newC^n$, by the induction hypothesis, either $\bv^n \rightsquigarrow \bw^k$ or $\bw^k \rightsquigarrow \bv^n$.
If $\bv^n \rightsquigarrow \bw^k$, then $\bv \rightsquigarrow \bw^k$, contradicting the assumption that $\bv || \bw^k$.
So \emph{it must be} that $\bw^k \rightsquigarrow \bv^n$.
But $\bv$ is a partial leader in $\graph(\bv^n)$, so by Lemma~\ref{lm:partialLeader}, $\bw^k$ cannot be a partial leader in $\graph(\bv^n)$ while being parallel to $\bv$.
This in turn implies (by Lemma~\ref{lm:alwaysLeader}) that $\bw^k$ is not a leader vertex.

Since $\bw^k$ is not a leader, it was added to $\newC$ from the past of another vertex $\bw^{k-1}$, which was added in round $k-1$ (observe that $\bw^k$ is a partial leader in $\graph(\bw^{k-1})$).
By the inductive hypothesis ($\newC^n$ is totally ordered by $\rightsquigarrow$), either $\bv^n \rightsquigarrow \bw^{k-1}$ or vice-versa.
If $\bv^n \rightsquigarrow \bw^{k-1}$, then $\graph(\bv^n) \subseteq \graph(\bw^{k-1})$.
Because $\bw^k$ is a partial leader in $\graph(\bw^{k-1})$, that would imply (by Lemma~\ref{lm:retroleader}) that $\bw^k$ is also a partial leader in $\graph(\bv^n)$, contradicting the fact that it isn't.
So \emph{it must be} that $\bw^{k-1} \rightsquigarrow \bv^n$ and that $\graph(\bw^{k-1}) \subseteq \graph(\bv^n)$.

Now, both $\bw^{k-1}$ and $\bv$ belong to $\graph(\bv^n)$, and $\bw^{k-1}$ cannot be parallel to $\bv$ ($k$ is the smallest round in which a vertex parallel to $\bv$ exists).
If $\bw^{k-1} \rightsquigarrow \bv$, then $\bw^k \rightsquigarrow \bv$, contradicting the assumption that $\bw^k || \bv$.
Let $\bv \rightsquigarrow \bw^{k-1}$.
Because $\bv$ is a partial leader in $\graph(\bv^n)$, by Lemma~\ref{lm:retroleader}, it is also a partial leader in $\graph(\bw^{k-1})$.
Therefore, both $\bw^k$ and $\bv$ are partial leaders in $\graph(\bw^{k-1})$.
By Lemma~\ref{lm:partialLeader}, either $\bv \rightsquigarrow \bw^k$ or vice-versa, which contradicts the assumption that $\bv || \bw^k$.

We conclude that no such vertex $\bw^k$ exists. Thus, for a round $k < n + 1$, there is no vertex $\bw'$ added to $\newC$ in round $k$ such that $\bw' || \bv$.

\subparagraph{Case (2).} Both $\bv$ and $\bw$ were added to $\newC$ in round $n+1$.
Let $\bv^n$ and $\bw^n$ be the vertices used to add $\bv$ and $\bw$ to $\newC$ respectively.
Then, $\bv^n$ and $\bw^n$ were added in round $n$, and $\bv$ is a partial leader in $\graph(\bv^n)$ (respectively, $\bw$ is a partial leader in $\graph(\bw^n)$).
By the inductive hypothesis, either $\bv^n \rightsquigarrow \bw^n$ or $\bw^n \rightsquigarrow \bv^n$ (since $\newC^n$ is totally ordered by $\rightsquigarrow$).
Suppose w.l.o.g. that $\bv^n \rightsquigarrow \bw^n$.
If $\bv^n \rightsquigarrow \bw$, then $\bv \rightsquigarrow \bw$, contradicting the assumption that $\bv||\bw$.
On the other hand, $\graph(\bv^n) \subseteq \graph(\bw^n)$.
So if $\bw \rightsquigarrow \bv^n$, Lemma~\ref{lm:retroleader} would imply that $\bw$ is a partial leader in $\graph(\bv^n)$ together with $\bv$, contradicting Lemma~\ref{lm:partialLeader}.
Therefore, $\bv^n || \bw$.
This is exactly Case (1), which leads to a contradiction.

\subparagraph{Conclusion.}
Assuming that there exist $\bv$ and $\bw$ in $\newC$ such that $\bv||\bw$ at the end of round $n+1$ leads to a contradiction in both cases (1) and (2).
We conclude that $\newC^{n+1}$ is totally ordered by $\rightsquigarrow$.
Since the value returned by $\getPreCommitted(G)$ is the final value of $\newC$, and $\newC$ is totally ordered by $\rightsquigarrow$ at the end of \emph{every round}, the set of pre-committed vertices is totally ordered by $\rightsquigarrow$.
\end{proof}
\end{lemma}

\begin{lemma}
\label{lm:includeC}
Let $\cC_i$ and $\cC_j$ be the sets of pre-committed vertices from $G_i$ and $G_j$ respectively. Then, $\cC_i \subseteq \cC_j \vee \cC_j \subseteq \cC_i$.

\begin{proof}
If either $\cC_i$ or $\cC_j$ is empty, the result is immediate.
Otherwise, let $\bv$ and $\bw$ be the last pre-committed vertices (with respect to $\rightsquigarrow$) in $\cC_i$ and $\cC_j$ respectively.
Since $\cC_i$ and $\cC_j$ are both totally ordered by $\rightsquigarrow$ (Lemma~\ref{lm:totalC}), $\bv$ and $\bw$ exist.
Moreover, both must be leaders in their respective graphs by construction.
Thus, either $\bv \rightsquigarrow \bw$ or $\bw \rightsquigarrow \bv$.
Assume w.l.o.g. that $\bv \rightsquigarrow \bw$ and suppose that there is a vertex $\bv' \in \cC_i$ satisfying: $\bv'$ is the \emph{last} vertex in $\cC_i$ (according to $\rightsquigarrow$) such that $\bv' \not\in \cC_j$.

By Lemma~\ref{lm:alwaysLeader}, if $\bv'$ is a leader in $G_i$, then it is a partial leader in $\graph(\bw)$.
Thus $\isPartial(\bv', \bw)$ holds and $\bv'$ should have been added to $\cC_j$ in the first while loop iteration of Algorithm~\ref{alg:preCommit} (line~\ref{ln:getPartials}).
If $\bv'$ is not a leader in $G_i$, then there is $\bv'' \in \cC_i$ for which $\isPartial(\bv', \bv'')$ holds, and $\bv'$ was added to $\cC_i$ at line~\ref{ln:getPartials} by using $\bv''$.
But $\bv'$ is the last vertex in $\cC_i$ that does not belong to $\cC_j$, so $\bv'' \in \cC_j$.
Consider the iteration in which $\bv''$ was added to $\cC_j$.
Then, in the following iteration, $\isPartial(\bv', \bv'')$ is true and $\bv'$ is added to $\cC_j$.

We conclude that there is no such vertex $\bv'$. Therefore, $\cC_i \subseteq \cC_j$.
\end{proof}
\end{lemma}

Now, we define $\cV_i$ as the permutation of $\cC_i$ such that $\bv$ precedes $\bw$ in $\cV_i$ iff $\bv \rightsquigarrow \bw$.

\begin{theorem}
\label{th:staticPrefix}
For all $\cV_i$ and $\cV_j$ it holds that $\cV_i \preceq \cV_j \vee \cV_j \preceq \cV_i$.
\begin{proof}
If any of $\cV_i$ or $\cV_j$ is an empty sequence, then the result is immediate.
Otherwise, let $\cC_i$ and $\cC_j$ be the set of pre-committed vertices of $G_i$ and $G_j$ respectively ($\cV_i$ is a permutation of $\cC_i$ and $\cV_j$ is a permutation of $\cC_j$).
By Lemma~\ref{lm:includeC}, either $\cC_i \subseteq \cC_j$ or vice-versa.
Assume w.l.o.g. that $\cC_i \subseteq \cC_j$.
Observe that every vertex that appears in $\cV_i$ also appears in $\cV_j$.
Let $\bv$ be the last vertex in $\cV_i$ (by construction, $\bv$ is a leader in $G_i$), and suppose that $\cV_i \not\preceq \cV_j$.
Then, there is a vertex in $\cV_j$ that precedes $\bv$ and it does not appear in $\cV_i$ (nor in $\cC_i$).
Let $\bw$ be the last vertex appearing in $\cV_j$ such that $\bw \rightsquigarrow \bv$ \emph{and} $\bw$ does not appear in $\cV_i$.
If $\bw$ is a leader vertex in $G_j$, then $\isPartial(\bw, \bv)$ holds and $\bw$ would have been added to $\cC_i$ at line~\ref{ln:getPartials} of Algorithm~\ref{alg:preCommit}.
This means that $\bw$ would appear in $\cV_i$, a contradiction.
Otherwise, there exists a vertex $\bw' \in \cC_j$ for which $\isPartial(\bw,\bw')$ holds, and $\bw'$ was used in line~\ref{ln:getPartials} to add $\bw$ to $\cC_j$.
Since $\bw$ is the last in $\cV_j$ to satisfy $\bw \rightsquigarrow \bv$ and $\bw \not\in \cC_i$, either $\bv \rightsquigarrow \bw'$ or $\bw' \in \cC_i$ (both cannot be true at the same time, since $\bv$ is the last vertex of $\cV_i$).
We consider each case in the following.

If $\bv \rightsquigarrow \bw'$, then $\graph(\bv) \subseteq \graph(\bw')$.
It follows from Lemma~\ref{lm:retroleader} that $\bw$ is a partial leader in $\graph(\bv)$, since it is a partial leader in $\graph(\bw')$.
Therefore, $\bw$ should have been added to $\cC_i$ at line~\ref{ln:getPartials} of Algorithm~\ref{alg:preCommit} ($\isPartial(\bw, \bv)$ is true), a contradiction.
Now, let $\bw' \in \cC_i$, and consider the iteration of the while loop (Algorithm~\ref{alg:preCommit}) in which $\bw'$ was added to $\cC_i$.
In the next iteration, $\isPartial(\bw, \bw')$ holds, and $\bw$ should have been added to $\cC_i$, a contradiction.

We conclude that no such vertex $\bw$ exists, therefore $\cV_i \preceq \cV_j$.

\end{proof}
\end{theorem}

\begin{theorem}
\label{th:monotonicity}
If $G_i$ is a subgraph of $G_j$, then $\cV_i \preceq \cV_j$.
\begin{proof}
If $\cV_i$ is empty, then the result is immediate.
Otherwise, let $\bv$ be the last vertex appearing in $\cV_i$. By construction, $\bv$ is a leader vertex in $G_i$, thus it is also a leader in $G_j$.
So $\bv \in \cC_j$ and $\bv$ appears in $\cV_j$.
Since the sequence $\cV_i$ ends at $\bv$, and because $\cV_i \preceq \cV_j \vee \cV_j \preceq \cV_i$ (Theorem~\ref{th:staticPrefix}), it follows that $\cV_i \preceq \cV_j$.
\end{proof}
\end{theorem}

The following liveness results hold in every infinite execution.

\begin{theorem}
\label{th:dagLive}
If a correct replica $r_i$ issues a vertex $\bv$, then $\bv$ eventually appears in the causal past of some vertex pre-committed by $r_i$.
\begin{proof}
From the communication component, every correct replica eventually receives $\bv$, and correct replicas periodically issue new vertices. Also, if a correct replica adds a vertex $\bw$ to its local DAG, then every correct replica eventually does so.
There exists an event $e_{\Omega}$ after which all queries output by $\Omega$ at correct replicas return the same correct replica $r_j$ (Eventual Accuracy and Eventual Agreement).
As a consequence, every new vertex issued by a correct replica has $r_j$ as a leader after $e_{\Omega}$.
Let $e_{\leader}$ be an event after $e_{\Omega}$ satisfying: the last vertex $r_j$ received from each correct replica has $r_j$ as its leader, and every subsequent vertex $r_j$ receives from correct replicas also has $r_j$ as its leader (note that such an event exists since replicas periodically issue new vertices).
Now, let $e_{\bv}$ be the event in which the last correct replica receives $\bv$, and let $e$ be the later of $e_{\leader}$ and $e_{\bv}$.
After $e$, $r_j$ eventually issues a new vertex $\bw$ for which $\bv \in \past(\bw)$ and there is a quorum of replicas in $\previous(\bw)$ that vote for $r_j$.
From the communication component, $\bw$ is eventually received and included in the past of a newly issued vertex from every correct replica.
Such vertices are then received by $r_i$, which gathers enough support from $\causalRegion(\bw)|_Q$ to pre-commit $\bw$, where $Q$ is a quorum containing only correct replicas, and the result follows.
\end{proof}
\end{theorem}

\subsection{Computing the Optimistic and Committed Sequences}
\label{subsec:computing}

Algorithm~\ref{alg:computing} describes the functions used to compute $\uncommittedS$ ($\getUncommitted$) and $\committedS$ ($\getCommitted$).
The algorithm repeatedly uses a topological sorting $f$ to produce a totally ordered sequence from an input graph.

In order to generate $\committedS_i$ and $\uncommittedS_i$, $r_i$ calls $\getCommitted(G_i)$ and $\getUncommitted(G_i)$ respectively.
The function $\getCommitted$ iterates through vertices from $\cV_i$ in order.
At each iteration, the next pre-committed vertex $\bv$ is used as a reference for sorting $\graph(\bv)$ with the topological sorting $f$, but only the vertices in $\graph(\bv)$ that are not yet committed are sorted, and the result is concatenated with the previously generated sequence.
The resulting sequence after iterating through all pre-committed vertices becomes the new committed sequence.

For $\getUncommitted$, on the other hand, only the uncommitted tail (vertices from $G_i$ that are not yet committed) is sorted using $f$, and the resulting sequence is concatenated with the committed prefix.

\begin{algorithm}
\begin{smartalgorithmic}[1]

    \Operation{\getCommitted}{$G$}
        \State $\preCommittedV \gets \cV(G)$ \label{line:preCommitted}
        \State $\newCommittedV \gets ()$
        \For{each $\bv \in \preCommittedV$ in sequence order}
            \State $\toCommitV \gets \present(\bv) - \present(\last(\newCommittedV))$
            \State $\newCommittedV \gets \newCommittedV + f(G|_{\toCommitV})$ \label{line:sorting}
        \EndFor
        \State \Return{$\newCommittedV$}
    \EndOperation

    \algspace[0.3]
    
    \Operation{\getUncommitted}{$G$}
        \State $\comV \gets \getCommitted(G)$
        \State $\uncTail \gets G.V - \{\comV\}$ \label{line:tail} \Comment{$\{\comV\}$ is the set of elements belonging to $\comV$.}
        \State \Return{$\comV + f(G|_{\uncTail})$} \label{line:inclusion}
    \EndOperation
    
    \algspace[0.3]

    \HelperFunctions
        \State $\last(\textsf{sequence})$: Returns the last element from \textsf{sequence} or $\perp$ if \textsf{sequence} = $()$.
        \State $\present(\bv)$: Returns $\past(\bv) \cup \{\bv\}$ or $\emptyset$ if $\bv = \perp$.
        \State $+$: Concatenation of sequences.
    \EndHelperFunctions

\end{smartalgorithmic}
\caption{Code for replica $r$ - Computation of sequences.}
\label{alg:computing}
\end{algorithm}

\begin{theorem}
\label{th:comSequence}
The sequences computed from $\getCommitted$ satisfy Prefix Consistency, Prefix Inclusion and Prefix Progress.
\begin{proof}
    The order on the reference vertices for commitment is defined by the function $\cV$ (line~\ref{line:preCommitted}), which, by Theorem~\ref{th:staticPrefix}, satisfies $\cV(G_i) \preceq \cV(G_j) \vee \cV(G_j) \preceq \cV(G_i)$ for all $G_i$ and $G_j$, implying Prefix Consistency.
    Prefix Inclusion is satisfied by construction (line~\ref{line:inclusion}).
    By Theorem~\ref{th:dagLive}, every vertex issued by a correct replica $r_i$ is eventually committed by $r_i$.
    Thus, if a vertex $\bv$ is received by $r_i$, $r_i$ will eventually issue a vertex with $\bv$ in its past and commit that vertex, which guarantees Prefix Progress.
\end{proof}
\end{theorem}

\begin{theorem}
\label{th:unComSequence}
The sequences computed from $\getUncommitted$ satisfy Strong Eventual-Consistency and Causal-Consistency.
\begin{proof}
    Strong Eventual-Consistency holds because every computation is deterministic within $\getUncommitted$.
    For Causal-Consistency, we must analyze both the committed and uncommitted sequences, since the former is a prefix of the latter (Theorem~\ref{th:comSequence}).

    By construction, the ordering of committed vertices preceding a reference (pre-committed) vertex follows the topological sorting $f$ (line~\ref{line:sorting}), which preserves the partial ordering of the graph.
    Also, for each new tail sequence that is committed, the vertices in the tail must either be parallel to the previous reference vertex, or come after it according to $\rightsquigarrow$.
    Therefore the committed sequence is consistent with the partial order determined by $\rightsquigarrow$.
    The uncommitted tail (sequence of vertices not yet committed) from line~\ref{line:tail} is also ordered according to $f$.
    Lastly, all vertices from the uncommitted tail are either parallel or come after the ones from the committed sequence, so the concatenation from line~\ref{line:inclusion} is consistent with $\rightsquigarrow$ and thus the algorithm satisfies Causal-Consistency.
\end{proof}
\end{theorem}

\subsection{Communication Complexity and Latency}
\label{subsec:complexity}

\subparagraph{Communication complexity.} Given a DAG produced by the communication component, Algorithm~\ref{alg:computing} derives both the optimistic and committed sequences using only local computation, and therefore incurs no additional communication overhead.
The communication complexity of the protocol is thus inherited from the underlying communication component, with the only additional cost arising from the implementation of the $\Omega$ oracle.

\subparagraph{Commitment latency.} We analyze the case of a stable leader replica, whose identifier is output by $\Omega$ at a majority of replicas for a sufficiently long period, as the commitment protocol cannot make progress otherwise.
Consider a vertex $\bv$ issued by a stable leader replica $r$, whose past already contains support from a quorum of replicas. The vertex can be committed at any replica within two message delays: one for $\bv$ to reach a quorum of replicas, and another for vertices from that quorum to reach all other replicas.
This measure, however, does not account for the time between receiving $\bv$ and issuing a subsequent vertex, which is application-dependent.

\section{Conclusion}
\label{sec:conclusion}

We presented a framework for optimistic replication based on causality DAGs.
In our approach, replicas execute operations using locally available information while exchanging messages asynchronously to ensure convergence.
By augmenting disseminated messages with the output of an eventual leader oracle, replicas are able to derive an efficient deterministic commitment directly from the causality DAG.
The resulting mechanism requires virtually no communication overhead beyond the dissemination layer and does not rely on rounds, randomization, or a separate ordering protocol.
The approach shows that causality DAGs, as well as PO-Logs, can serve as practical constructions for commitment in optimistically replicated systems.

\bibliography{main}

\end{document}

%% file: definitions.tex
\NewDeclarationType{DistributedObject}{\textsc}
\NewDeclarationType{Function}{\textsf}
\NewDeclarationType{Atribute}{\textsf}
\NewDeclarationType{Event}{\textsf}
\NewDeclarationType{Operation}{\textbf}
\NewDeclarationType{Variable}{\mathit}
\NewDeclarationType{Property}{\textrm}
\NewDeclarationType{Type}{\textrm}
\NewDeclarationType{MessageType}{\textbf}
\NewDeclarationType{Constant}{\textbf}
\NewDeclarationType{Term}{}

\mathchardef\mhyphen="2D

\newcommand{\Message}[1]{\ensuremath{\langle}#1\ensuremath{\rangle}}

\algblockdefx[Multiline]{Multiline}{EndMultiline}{}{}
\algtext*{EndMultiline}  

\algblockdefx[Function]{Function}{EndFunction}[2]{\textbf{function} #1(#2)}{}
\algtext*{EndFunction}  

\algblockdefx[ForEach]{ForEach}{EndForEach}[1]{\textbf{for each} #1 \textbf{do}}{}
\algtext*{EndForEach}  

\algblockdefx[Repeat]{Repeat}{EndRepeat}[1]{\textbf{repeat} #1}{}
\algtext*{EndRepeat}

\algblockdefx[Procedure]{Procedure}{EndProcedure}[2]{\textbf{procedure} #1(#2)}{}
\algtext*{EndProcedure}  

\algblockdefx[Operation]{Operation}{EndOperation}[2]{\textbf{operation} #1(#2)}{}
\algtext*{EndOperation}  

\algblockdefx[OverrideOperation]{OverrideOperation}{EndOperation}[2]{\textbf{override operation} #1(#2)}{}
\algtext*{EndOperation}  

\algblockdefx[Upon]{Upon}{EndUpon}[1]{\textbf{upon} #1}{}
\algtext*{EndUpon}  

\algblockdefx[Behavior]{Behavior}{EndBehavior}[1]{\textbf{behavior} of #1}{}
\algtext*{EndBehavior}  

\algblockdefx[UponEvent]{UponEvent}{EndUponEvent}[1]{\textbf{upon event} #1}{}
\algtext*{EndUponEvent}  

\algblockdefx[UponReceiving]{UponReceiving}{EndUponReceiving}[2]{\textbf{upon} \Receive~\Message{#1} \textbf{from} #2}{}
\algtext*{EndUponReceiving}  

\algblockdefx[UponConsistentDeliver]{UponConsistentDeliver}{EndUponConsistentDeliver}[2]{\textbf{upon consistently delivering} $\langle$#1$\rangle$ \textbf{from} #2}{}
\algtext*{EndUponConsistentDeliver}  

\algblockdefx[ProcessState]{ProcessState}{EndProcessState}{\textbf{State:}}{}
\algtext*{EndProcessState}  

\algblockdefx[DistributedObjects]{DistributedObjects}{EndDistributedObjects}{\textbf{Distributed objects:}}{}
\algtext*{EndDistributedObjects}  

\algblockdefx[BehaviorLA]{BehaviorLA}{EndBehaviorLA}{\textbf{Behavior upon receiving a message $m$ of instance $LA_k$:}}{}
\algtext*{EndBehaviorLA}  

\algblockdefx[FunctionDeclaration]{FunctionDeclaration}{EndFunctionDeclaration}{\textbf{Functions:}}{}
\algtext*{EndFunctionDeclaration}

\algblockdefx[StaticLocalVariables]{StaticLocalVariables}{EndStaticLocalVariables}{\textbf{Static Local Variables:}}{}
\algtext*{EndStaticLocalVariables}  

\algblockdefx[HelperFunctions]{HelperFunctions}{EndHelperFunctions}{\textbf{Helper functions:}}{}
\algtext*{EndHelperFunctions}  

\algnewcommand{\Returns}[1]{\textbf{returns} {#1}}

\algnewcommand{\Break}{\textbf{break}}

\algnewcommand{\WaitFor}{\textbf{wait for} }
\algnewcommand{\WaitUntil}{\textbf{wait until} }
\algnewcommand{\WaitReceived}[1]{\textbf{wait until } #1 \textbf{ is received}}
\algnewcommand{\Send}[2]{\textbf{send} $\langle$#1$\rangle$ \textbf{to} #2}
\algnewcommand{\IfThenElse}[3]{
    \algorithmicif\ #1\ \algorithmicthen\ #2\ \algorithmicelse\ #3}
\algnewcommand{\IfThen}[2]{
    \algorithmicif\ #1\ \algorithmicthen\ #2}

\algnewcommand{\TriggerEvent}[1]{\textbf{trigger event }#1}

\algnewcommand{\Parameters}[1]{\State \textbf{Instance parameters: }#1}

\DeclareGlobalConstant{false}
\DeclareGlobalConstant{true}
\DeclareGlobalConstant{none}

\DeclareGlobalEvent{Receive}
\DeclareGlobalEvent{Learn}
\DeclareGlobalEvent{Startup}

\DeclareGlobalDistributedObject[rdag]{D-DAG}
\DeclareGlobalDistributedObject{DAG}

\DeclareGlobalAtribute{op}
\DeclareGlobalAtribute{rp}
\DeclareGlobalAtribute{req}
\DeclareGlobalAtribute{seq}
\DeclareGlobalAtribute{m}
\DeclareGlobalAtribute{id}
\DeclareGlobalAtribute{leader}
\DeclareGlobalAtribute{data}
\DeclareGlobalAtribute{manner}

\DeclareGlobalFunction[getPreCommitted]{get\_pre\_committed}
\DeclareGlobalFunction{disseminate}
\DeclareGlobalFunction{join}
\DeclareGlobalFunction{remove}
\DeclareGlobalFunction{requestOp}
\DeclareGlobalFunction{getStableMembers}
\DeclareGlobalFunction{getUnstableMembers}
\DeclareGlobalFunction{apply}
\DeclareGlobalFunction{query}
\DeclareGlobalFunction[getCommitted]{get\_committed}
\DeclareGlobalFunction[getUncommitted]{get\_uncommitted}
\DeclareGlobalFunction[getPrefix]{get\_prefix}
\DeclareGlobalFunction{minimal}
\DeclareGlobalFunction{last}
\DeclareGlobalFunction[nextToCommit]{next\_to\_commit}
\DeclareGlobalFunction[newMembers]{new\_members}
\DeclareGlobalFunction{present}

\newcommand{\bv}{\textbf{v}}
\newcommand{\bw}{\textbf{w}}
\newcommand{\bz}{\textbf{z}}

\newcommand{\past}{\textit{past}}
\newcommand{\nnext}{\textit{next}}
\newcommand{\previous}{\textit{previous}}
\newcommand{\current}{\textit{current}}
\newcommand{\rep}{\textit{rep}}
\newcommand{\votes}{\textit{votes}}
\newcommand{\trace}{\textit{trace}}
\newcommand{\causalRegion}{\textit{causal-region}}

\newcommand{\graph}{\textit{graph}}
\newcommand{\isLeader}{\textit{is\_leader}}
\newcommand{\isPartial}{\textit{is\_partial}}

\DeclareGlobalVariable{oldC}
\DeclareGlobalVariable{newC}

\DeclareGlobalVariable{unstableM}
\DeclareGlobalVariable{stableM}
\DeclareGlobalVariable[view]{View}
\DeclareGlobalVariable{unstableV}
\DeclareGlobalVariable{stableV}
\DeclareGlobalVariable[hist]{Hist}
\DeclareGlobalVariable{uncommittedS}
\DeclareGlobalVariable{committedS}
\DeclareGlobalVariable{uncommittedM}
\DeclareGlobalVariable{committedM}
\DeclareGlobalVariable[toNotify]{to\_notify}
\DeclareGlobalVariable{uncV}
\DeclareGlobalVariable{comV}
\DeclareGlobalVariable{uncTail}
\DeclareGlobalVariable[preCommittedV]{pre\_committedV}
\DeclareGlobalVariable[newCommittedV]{new\_committedV}
\DeclareGlobalVariable[toCommitV]{to\_commitV}
\DeclareGlobalVariable{members}

\newcommand{\prev}{\textit{prev}}

\newcommand{\llast}{\textit{last}}

\newcommand{\cT}{\mathcal{T}}

\newcommand{\cV}{\mathcal{V}}
\newcommand{\cC}{\mathcal{C}}

\newcommand{\cA}{\mathcal{A}}

\DeclareGlobalMessageType{REQUEST}
\DeclareGlobalMessageType[WREPLY]{WEAK-REPLY}
\DeclareGlobalMessageType[SREPLY]{STRONG-REPLY}

\newcommand{\REVIEW}{1} 
\ifnum\REVIEW=1
    \newcommand{\petr}[1]{{\color{red}{[Petr: #1]}}}
    \newcommand{\sara}[1]{{\color{blue!50!black}{[Sara: #1]}}}
    \newcommand{\joao}[1]{{\color{green!50!black}{[Joao: #1]}}}
\else
    \newcommand{\joao}[1]{}
    \newcommand{\petr}[1]{}
    \newcommand{\sara}[1]{}
\fi

%% file: definitions_tickz.tex
\definecolor{causalRegionColor}{HTML}{557fae}
\definecolor{previousColor}{HTML}{0057B8}
\definecolor{nextColor}{HTML}{B21F2D}
\definecolor{leaderSupportColor}{HTML}{6cdbbe}

\pgfdeclareradialshading{support radial}{\pgfpointorigin}{
    color(0bp)=(white);
    color(4bp)=(leaderSupportColor!15!white);
    color(25bp)=(leaderSupportColor!30!white)
}

\newcommand{\votevalue}[1]{\mathbf{#1}}

\newcommand{\legendvotevertex}{%
    \begin{tabular}{@{}c@{}}
        \tikz[baseline=-0.2ex]\node[
            font=\scriptsize,
            circle,
            draw=black,
            line width=0.5pt,
            minimum size=9pt,
            inner sep=0.3pt
        ] {$\votevalue{q}$};\\[-2pt]
        $\bv_{p:n}$
    \end{tabular}%
}

\newcommand{\legendleadervertex}{%
    \begin{tabular}{@{}c@{}}
        \tikz[baseline=-0.2ex]\node[
            font=\scriptsize,
            rectangle,
            draw=black,
            line width=0.75pt,
            minimum size=9pt,
            inner sep=0.3pt
        ] {$\votevalue{q}$};\\[-2pt]
        $\bv_{p:n}$
    \end{tabular}%
}

\newcommand{\causalregionswatch}{%
    \tikz[baseline=0ex]{
        \path[
            draw=none,
            fill=causalRegionColor!16,
            fill opacity=0.75,
            rounded corners=2pt,
            blur shadow={shadow blur radius=2pt,shadow blur steps=8,
                shadow xshift=0pt,shadow yshift=0pt,shadow opacity=12}
        ] (0,0) rectangle (0.7,0.3);
    }%
}

\newcommand{\previousswatch}{%
    \tikz[baseline=0ex]{
        \path[
            pattern={Lines[angle=-45,distance=3pt,line width=0.6pt]},
            pattern color=previousColor!52
        ] (0,0) rectangle (0.7,0.3);
    }%
}

\newcommand{\nextswatch}{%
    \tikz[baseline=0ex]{
        \path[
            pattern={Lines[angle=45,distance=3pt,line width=0.6pt]},
            pattern color=nextColor!52
        ] (0,0) rectangle (0.7,0.3);
    }%
}

\newcommand{\supportsetswatch}{%
    \tikz[baseline=0ex]{
        \begin{scope}
            \clip[rounded corners=2pt] (0,0) rectangle (0.7,0.3);
            \shade[
                shading=support radial
            ] (0.35,0.15) circle[radius=0.5cm];
        \end{scope}
        \draw[draw=leaderSupportColor,line width=0.5pt,rounded corners=2pt]
            (0,0) rectangle (0.7,0.3);
    }%
}

\newcommand{\causallegend}{%
    \begingroup
    \small
    \setlength{\tabcolsep}{2pt}%
    \renewcommand{\arraystretch}{1.25}%
    \begin{tabular}{@{}cl@{\hspace{1.6em}}cl@{}}
        $\dashrightarrow$ & Causal order
            & \causalregionswatch & Causal region of leader $\bv$ \\
        $\rightarrow$ & Same-process order
            & \previousswatch & $\mathit{previous}(\bv)$ \\
        \legendvotevertex & Vertex $n$ from process $p$, voting for $q$
            & \nextswatch & $\mathit{next}(\bv)$ \\
        \legendleadervertex & Leader or Partial Leader vertex
            & \supportsetswatch & Set of vertices supporting $\bv$
    \end{tabular}%
    \endgroup
}

\newcommand{\replicarows}{%
    \foreach \replica/\row in {a/3,b/2,c/1,d/0} {
        \node[anchor=east,inner sep=0pt] at (-0.9,\row)
            {Replica $\replica$};
    }
    \pgfresetboundingbox
    \path[use as bounding box] (-2.5,-0.55) rectangle (8.7,3.8);
}

%% file: main.bib
@article{lamport1978time,
  title={Time, Clocks, and the Ordering of Events in a Distributed System},
  author={Lamport, Leslie},
  journal={Communications},
  year={1978}
}

@book{gutin2009digraphs,
  title={Digraphs: Theory, Algorithms and Applications},
  author={Gutin, Gregory Z and Bang-Jensen, J{\o}rgen},
  year={2009},
  publisher={Springer-Verlag London}
}

@article{chandra1996weakest,
  title={The weakest failure detector for solving consensus},
  author={Chandra, Tushar Deepak and Hadzilacos, Vassos and Toueg, Sam},
  journal={Journal of the ACM (JACM)},
  volume={43},
  number={4},
  pages={685--722},
  year={1996},
  publisher={ACM New York, NY, USA}
}

@article{burckhardt2014replicated,
  title={Replicated data types: specification, verification, optimality},
  author={Burckhardt, Sebastian and Gotsman, Alexey and Yang, Hongseok and Zawirski, Marek},
  journal={ACM Sigplan Notices},
  volume={49},
  number={1},
  pages={271--284},
  year={2014},
  publisher={ACM New York, NY, USA}
}

@article{almeida2024approaches,
  title={Approaches to conflict-free replicated data types},
  author={Almeida, Paulo S{\'e}rgio},
  journal={ACM Computing Surveys},
  volume={57},
  number={2},
  pages={1--36},
  year={2024},
  publisher={ACM New York, NY}
}

@inproceedings{keidar2021all,
  title={All you need is dag},
  author={Keidar, Idit and Kokoris-Kogias, Eleftherios and Naor, Oded and Spiegelman, Alexander},
  booktitle={Proceedings of the 2021 ACM Symposium on Principles of Distributed Computing},
  pages={165--175},
  year={2021}
}

@inproceedings{danezis2022narwhal,
  title={Narwhal and tusk: a dag-based mempool and efficient bft consensus},
  author={Danezis, George and Kokoris-Kogias, Lefteris and Sonnino, Alberto and Spiegelman, Alexander},
  booktitle={Proceedings of the Seventeenth European Conference on Computer Systems},
  pages={34--50},
  year={2022}
}

@article{moser1999byzantine,
  title={Byzantine-resistant total ordering algorithms},
  author={Moser, Louise E and Melliar-Smith, Peter M},
  journal={Information and Computation},
  volume={150},
  number={1},
  pages={75--111},
  year={1999},
  publisher={Elsevier}
}

@article{baird2016swirlds,
  title={The swirlds hashgraph consensus algorithm: Fair, fast, byzantine fault tolerance},
  author={Baird, Leemon},
  journal={Swirlds Tech Reports SWIRLDS-TR-2016-01, Tech. Rep},
  volume={34},
  pages={9--11},
  year={2016}
}

@inproceedings{gkagol2019aleph,
  title={Aleph: Efficient atomic broadcast in asynchronous networks with byzantine nodes},
  author={G{\k{a}}gol, Adam and Le{\'s}niak, Damian and Straszak, Damian and {\'S}wi{\k{e}}tek, Micha{\l}},
  booktitle={Proceedings of the 1st ACM Conference on Advances in Financial Technologies},
  pages={214--228},
  year={2019}
}

@inproceedings{keidar2023cordial,
  title={Cordial Miners: Fast and Efficient Consensus for Every Eventuality},
  author={Keidar, Idit and Naor, Oded and Poupko, Ouri and Shapiro, Ehud},
  booktitle={37th International Symposium on Distributed Computing},
  year={2023}
}

@article{babel2023mysticeti,
  title={Mysticeti: Reaching the limits of latency with uncertified dags},
  author={Babel, Kushal and Chursin, Andrey and Danezis, George and Kichidis, Anastasios and Kokoris-Kogias, Lefteris and Koshy, Arun and Sonnino, Alberto and Tian, Mingwei},
  journal={arXiv preprint arXiv:2310.14821},
  year={2023}
}

@book{cachin2011introduction,
  title={Introduction to reliable and secure distributed programming},
  author={Cachin, Christian and Guerraoui, Rachid and Rodrigues, Lu{\'\i}s},
  year={2011},
  publisher={Springer Science \& Business Media}
}

@article{waitfree2025,
       author = {{Kuznetsov}, Petr and {Perion}, Maxence and {Tucci-Piergiovanni}, Sara},
        title = "{Wait-free Replicated Data Types and Fair Reconciliation}",
      journal = {arXiv e-prints},
         year = 2025,
        month = aug,
          eid = {arXiv:2508.18193},
        pages = {arXiv:2508.18193},
          doi = {10.48550/arXiv.2508.18193},
archivePrefix = {arXiv},
       eprint = {2508.18193},
 primaryClass = {cs.DC},
       adsurl = {https://ui.adsabs.harvard.edu/abs/2025arXiv250818193K}
}

@article{saito2005optimistic,
  title={Optimistic replication},
  author={Saito, Yasushi and Shapiro, Marc},
  journal={ACM Computing Surveys (CSUR)},
  volume={37},
  number={1},
  pages={42--81},
  year={2005},
  publisher={ACM New York, NY, USA}
}

@article{cassandra,
  author       = {Avinash Lakshman and
                  Prashant Malik},
  title        = {Cassandra: a decentralized structured storage system},
  journal      = {{ACM} {SIGOPS} Oper. Syst. Rev.},
  volume       = {44},
  number       = {2},
  pages        = {35--40},
  year         = {2010}
}

@inproceedings{bayou,
  author       = {Douglas B. Terry and
                  Marvin Theimer and
                  Karin Petersen and
                  Alan J. Demers and
                  Mike Spreitzer and
                  Carl Hauser},
  title        = {Managing Update Conflicts in Bayou, a Weakly Connected Replicated
                  Storage System},
  booktitle    = {{SOSP}},
  year         = {1995}
}

@article{vogels-ec,
  author       = {Werner Vogels},
  title        = {Eventually consistent},
  journal      = {Commun. {ACM}},
  volume       = {52},
  number       = {1},
  pages        = {40--44},
  year         = {2009}
}

@inproceedings{dynamo,
  author       = {Giuseppe DeCandia and
                  Deniz Hastorun and
                  Madan Jampani and
                  Gunavardhan Kakulapati and
                  Avinash Lakshman and
                  Alex Pilchin and
                  Swaminathan Sivasubramanian and
                  Peter Vosshall and
                  Werner Vogels},
  title        = {Dynamo: amazon's highly available key-value store},
  booktitle    = {{SOSP}},
  pages        = {205--220},
  year         = {2007}
}

@inproceedings{spiegelman2022bullshark,
  title={Bullshark: Dag bft protocols made practical},
  author={Spiegelman, Alexander and Giridharan, Neil and Sonnino, Alberto and Kokoris-Kogias, Lefteris},
  booktitle={Proceedings of the 2022 ACM SIGSAC Conference on Computer and Communications Security},
  pages={2705--2718},
  year={2022}
}

@inproceedings{amores2024we,
  title={We will DAG you},
  author={Amores-Sesar, Ignacio and Cachin, Christian},
  booktitle={European Symposium on Research in Computer Security},
  pages={276--291},
  year={2024},
  organization={Springer}
}

@inproceedings{jovanovic2025mahi,
  title={Mahi-mahi: Low-latency asynchronous bft dag-based consensus},
  author={Jovanovic, Philipp and Kokoris-Kogias, Lefteris and Kumara, Bryan and Sonnino, Alberto and Tennage, Pasindu and Zablotchi, Igor},
  booktitle={2025 IEEE 45th International Conference on Distributed Computing Systems (ICDCS)},
  pages={549--559},
  year={2025},
  organization={IEEE}
}

@inproceedings{shapiro2011conflict,
  title={Conflict-free replicated data types},
  author={Shapiro, Marc and Pregui{\c{c}}a, Nuno and Baquero, Carlos and Zawirski, Marek},
  booktitle={Symposium on Self-Stabilizing Systems},
  pages={386--400},
  year={2011},
  organization={Springer}
}

@article{baquero2017pure,
  title={Pure operation-based replicated data types},
  author={Baquero, Carlos and Almeida, Paulo S{\'e}rgio and Shoker, Ali},
  journal={arXiv preprint arXiv:1710.04469},
  year={2017}
}

@article{ladin1992providing,
  title={Providing high availability using lazy replication},
  author={Ladin, Rivka and Liskov, Barbara and Shrira, Liuba and Ghemawat, Sanjay},
  journal={ACM Transactions on Computer Systems (TOCS)},
  volume={10},
  number={4},
  pages={360--391},
  year={1992},
  publisher={ACM New York, NY, USA}
}

@inproceedings{lloyd2011don,
  title={Don't settle for eventual: scalable causal consistency for wide-area storage with COPS},
  author={Lloyd, Wyatt and Freedman, Michael J and Kaminsky, Michael and Andersen, David G},
  booktitle={Proceedings of the Twenty-Third ACM Symposium on Operating Systems Principles},
  pages={401--416},
  year={2011}
}

@inproceedings{guerraoui2016incremental,
  title={Incremental consistency guarantees for replicated objects},
  author={Guerraoui, Rachid and Pavlovic, Matej and Seredinschi, Dragos-Adrian},
  booktitle={12th USENIX Symposium on Operating Systems Design and Implementation (OSDI 16)},
  pages={169--184},
  year={2016}
}

@inproceedings{li2012making,
  title={Making $\{$Geo-Replicated$\}$ systems fast as possible, consistent when necessary},
  author={Li, Cheng and Porto, Daniel and Clement, Allen and Gehrke, Johannes and Pregui{\c{c}}a, Nuno and Rodrigues, Rodrigo},
  booktitle={10th USENIX Symposium on Operating Systems Design and Implementation (OSDI 12)},
  pages={265--278},
  year={2012}
}

@inproceedings{li2018fine,
  title={Fine-grained consistency for geo-replicated systems},
  author={Li, Cheng and Pregui{\c{c}}a, Nuno and Rodrigues, Rodrigo},
  booktitle={2018 USENIX Annual Technical Conference (USENIX ATC 18)},
  pages={359--372},
  year={2018}
}

@article{kokocinski2021mixing,
  title={On mixing eventual and strong consistency: acute cloud types},
  author={Kokoci{\'n}ski, Maciej and Kobus, Tadeusz and Wojciechowski, Pawe{\l} T},
  journal={IEEE Transactions on Parallel and Distributed Systems},
  volume={33},
  number={6},
  pages={1338--1356},
  year={2021},
  publisher={IEEE}
}

@article{wojciechowski2026creek,
  title={Creek: A Single-Order Mixed-Consistency Replication Scheme},
  author={Wojciechowski, Pawe{\l} T and Kokoci{\'n}ski, Maciej and Kobus, Tadeusz},
  journal={Theoretical Computer Science},
  pages={115927},
  year={2026},
  publisher={Elsevier}
}

@inproceedings{aguilera2001stable,
  title={Stable leader election},
  author={Aguilera, Marcos K and Delporte-Gallet, Carole and Fauconnier, Hugues and Toueg, Sam},
  booktitle={International Symposium on Distributed Computing},
  pages={108--122},
  year={2001},
  organization={Springer}
}

@inproceedings{schiper2008robust,
  title={A robust and lightweight stable leader election service for dynamic systems},
  author={Schiper, Nicolas and Toueg, Sam},
  booktitle={2008 IEEE International Conference on Dependable Systems and Networks With FTCS and DCC (DSN)},
  pages={207--216},
  year={2008},
  organization={IEEE}
}

@inproceedings{dubois2015weakest,
  title={The weakest failure detector for eventual consistency},
  author={Dubois, Swan and Guerraoui, Rachid and Kuznetsov, Petr and Petit, Franck and Sens, Pierre},
  booktitle={Proceedings of the 2015 ACM Symposium on Principles of Distributed Computing},
  pages={375--384},
  year={2015}
}
